\documentclass[11pt]{article}
	
	\newcommand{\blind}{0}
	
    \makeatletter
    
    \renewcommand\section{\@startsection{section}{1}{\z@}%
                                       {-3.5ex \@plus -1ex \@minus -.2ex}%
                                       {2.3ex \@plus.2ex}%
                                       {\normalfont\normalsize\fontfamily{phv}\fontsize{16}{19}\bfseries}}
    \renewcommand\subsection{\@startsection{subsection}{2}{\z@}%
                                         {-3.25ex\@plus -1ex \@minus -.2ex}%
                                         {1.5ex \@plus .2ex}%
                                         {\normalfont\fontfamily{phv}\fontsize{14}{17}\bfseries}}
    \renewcommand\subsubsection{\@startsection{subsubsection}{3}{\z@}%
                                        {-3.25ex\@plus -1ex \@minus -.2ex}%
                                         {1.5ex \@plus .2ex}%
                                         {\normalfont\normalsize\fontfamily{phv}\fontsize{12}{17}\selectfont}}
    \makeatother
	\usepackage{amsmath}
	\usepackage{graphicx}
    \usepackage{amsmath}
    \usepackage{amssymb}
	\usepackage{enumerate}
    \usepackage{orcidlink}
    \usepackage{enumitem}
	\usepackage{xcolor}
    \usepackage{booktabs}
    \usepackage{tabularx}
    \usepackage{longtable}
    \usepackage{eurosym}
    \usepackage{arydshln}
    \usepackage{array}
    \usepackage[T1]{fontenc}
    \usepackage{booktabs}
    \usepackage[utf8]{inputenc} 
    \usepackage{lmodern}
    \usepackage{subcaption}
    \usepackage{textcomp}
    \usepackage{xcolor}
    \usepackage{float}
    \usepackage{threeparttable} 
    \usepackage{indentfirst}
	
	\usepackage{url} 
    \usepackage{algorithm}
    \usepackage[noend]{algpseudocode}
    \usepackage[super]{nth}
    \usepackage{hyperref}
    \usepackage{flafter}   
    \usepackage{placeins}
    \usepackage{needspace}   
    \usepackage{siunitx}
    \usepackage{graphicx}
    \usepackage{color,soul}
	\usepackage{todonotes}
    \usepackage[T1]{fontenc}
    \usepackage[left=0.7in,right=0.7in,top=0.75in,bottom=0.75in]{geometry}
    \usepackage[utf8]{inputenc}
	\usepackage{amsfonts, amsthm, latexsym, amssymb}
	\usepackage{lipsum} 
    \usepackage{xcolor}
    \usepackage{soul}
    \usepackage{orcidlink}  
    \usepackage{amssymb} 

    \colorlet{donecolor}{green!25}
    
    \usepackage{tablefootnote}
	\newcolumntype{L}[1]{>{\raggedright\arraybackslash}p{#1}}
    \newcolumntype{C}[1]{>{\centering\arraybackslash}p{#1}}
    \usepackage[super,comma,sort&compress]{natbib} 
\begin{document}

		\def\spacingset#1{\renewcommand{\baselinestretch}%
			{#1}\small\normalsize} \spacingset{1}
        \if0\blind
		{
			\title{\bf Brittle by Design? Multi-Scenario Reinforcement Learning for Robust Mission Coordination in Systems of Systems 
   }

            \author{İbrahim Oğuz Çetinkaya$^1$\,\orcidlink{0009-0008-8354-5968}, Sajad Khodadadian$^1$\,\orcidlink{0000-0002-5197-4652} and Taylan G. Topcu$^1$\,\orcidlink{0000-0002-0110-312X} \\
            $^1$\small Grado Department of Industrial and Systems Engineering, Virginia Tech, Blacksburg, VA }
			\maketitle
	     }  \fi

		\if1\blind
		{

            \title{\bf \emph{IISE Transactions} \LaTeX \ Template}
			\author{Author information is purposely removed for double-blind review}
			
\bigskip
			\bigskip
			\bigskip
			\begin{center}
				{\LARGE\bf \emph{IISE Transactions} \LaTeX \ Template}
			\end{center}
			\medskip
		} \fi
		\bigskip

	\spacingset{1.5} 
\vspace{-1cm}

\begin{abstract}
As systems engineering (SE) shifts from monolithic systems to Systems of Systems, Mission Engineering (ME) has emerged as a new line of thinking in SE. Mission environments are uncertain and dynamic, and outcomes depend on how mission assets interact with them, rendering static architectures brittle and calling for flexible yet analytically rigorous ME approaches. We propose an intelligent mission coordination methodology integrating high-fidelity digital mission models with Reinforcement Learning (RL) for adaptive task allocation. Using a Digital Engineering (DE) infrastructure comprising a digital mission model and agent-based simulation, we formulate mission tactics management as a Markov Decision Process and train an RL coordinator with Proximal Policy Optimization to map mission states to tactical actions from realized outcomes. An aerial firefighting mission with three scenarios demonstrates the approach. First, all trained coordinators significantly outperform random and heuristic baselines. Second, single-scenario coordinators can be brittle: the coordinator trained on the more forgiving scenario degrades to random tactics performance in more demanding scenarios, whereas the demanding-scenario coordinator retains most of its advantage when transferred. Third, within the same training budget, a coordinator trained on both scenarios concurrently matches the scenario-specific coordinators in their own scenarios and significantly outperforms every other policy. Sensitivity analyses on ignition location, weather, scenario, and training curricula show that the variability represented during training shapes post-deployment performance. This proof of concept shows DE-enabled mission simulations combined with RL offer a reusable methodology for improving ME practice, which can be extended to fleet design and selection from a mission-first perspective.

\textbf{Keywords:} Mission Engineering, System of Systems, Reinforcement Learning, Disaster Management, Artificial Intelligence for Systems Engineering (AI4SE), Robustness

\end{abstract}

\section{Introduction} 
\label{s:intro}

Systems Engineering (SE) has traditionally focused on the realization and lifecycle management of complex engineered systems within a given boundary ~\cite{Kossiakoff2011}. However, many modern capabilities are delivered by Systems of Systems (SoS): collections of operationally and managerially independent systems that together achieve what none can achieve alone ~\cite{maier1998}. Research on SoS architectures has made substantial progress in application areas such as aerospace~\cite{oneill2010, selva2017, grogan2016}, disaster management~\cite{fan2020}, supply chains~\cite{choi2018, jaradat2017}, and healthcare~\cite{mogahed2025, wickramasinghe2007healthcare, mogahed2026ontology}, among many others~\cite{delaurentis2005, raz2020, sosworkbench}. Nevertheless, deployed SoS often remain brittle. They are costly to modify once fielded, and their performance is sensitive to operational and contextual changes~\cite{raz2024ME, sapol2022}. Ensuring that an SoS remains effective therefore requires a perspective anchored in the missions it serves and extending to how its constituent systems are coordinated during operations.

To address this need, Mission Engineering (ME) has been proposed as an interdisciplinary ``process'' that bridges the gap between a mission and the engineering activities that support it; spanning the analysis, design, and integration of operational needs and capabilities toward desired mission outcomes~\cite{dod2020ME}. Within ME, two tightly coupled decision problems arise, reflecting needs identified by Raz et al.~\cite{raz2024ME}: selecting and configuring the assets that form the SoS for a mission, and managing those assets during the mission by adaptively allocating tasks as conditions unfold. 

Coordinating assets during mission execution is challenging due to unpredictably evolving threats and environmental conditions, nonlinear asset interactions, and a combinatorially growing space of possible task assignments~\cite{dachowicz_mission_2021, hazelrigg_framework_1998}. These factors often make mission decision problems intractable for exact methods such as optimization, and even when a tractable formulation exists, solving it at every decision point is too slow for operational timelines \cite{mosleh2016}. Overcoming these limitations requires two complementary capabilities: (i) a realistic environment in which coordination policies can be developed and tested and (ii) a decision method fast enough for operational timelines. Digital Engineering (DE) offers the former. Defined as an approach that aligns data-driven infrastructure, methods, innovation, and workforce to deliver high-payoff solutions in a timely manner~\cite{henderson2023de, de4me}, DE relies on practices such as model-based systems engineering, high-fidelity simulations, and digital twins. These practices could supply rich mission data and the infrastructure to address core ME needs via analytical tools~\cite{lee2021digital, pennock2022enabling}, and they could serve as a testbed for evaluating ME policies against realistic scenarios before the SoS is deployed. Reinforcement Learning (RL) provides the latter capability. An agent could learn a coordination policy offline by interacting with a mission simulation and, once trained, could map the current mission state to actions without solving the problem at each decision point~\cite{sutton1999}. 

Such learning-based coordination is gaining momentum, from hierarchical RL for SoS governance ~\cite{chen2022dynamic, chen2024sosconductor} to RL for defense and space mission planning~\cite{RL4UAV, RL4spaceplanning}. Nonetheless, most of these studies learn in abstract or simplified environments, and those that do use high-fidelity simulations do not frame their problems in ME terms. Hence, coupling RL with high-fidelity mission simulations for ME, so that coordination policies are learned through realistic mission dynamics, remains largely unexplored.

This paper addresses this gap by proposing an analytical methodology that integrates high-fidelity digital mission models with RL for intelligent mission coordination. Because a coordinator trained in one mission scenario may not perform well in another, we also examine how the scenarios used for training shape coordination performance. Specifically, we tackle three research questions:
\begin{enumerate}[label=\textbf{RQ\arabic*:}, leftmargin=*]
    \item How can RL methods be synergistically integrated with high-fidelity simulation models to enable intelligent mission coordination given uncertainties, and do the resulting coordinators outperform benchmark approaches?
    \item How brittle is the performance of intelligent mission coordination agents trained on a single scenario when they face new scenarios within the same mission?
    \item How does multi-scenario training compare with scenario-specific training in terms of overall performance and robustness across scenarios?
\end{enumerate}

\indent These research questions directly address the second ME need Raz pointed out: ability to adaptively allocate tasks and reconfigure the SoS \cite{raz2024ME}. Throughout, we use \emph{brittleness} to denote the loss of coordination performance when a coordinator is deployed under conditions it did not encounter during training, and \emph{robustness} to denote the retention of that performance. Mission resilience in the broader SE sense, the ability of the SoS to absorb, adapt to, and recover from disruptions that occur during a mission, is the longer-term objective of this research agenda; we return to the distinction in Section~\ref{sec:discussion}. 

Building on our previous studies \cite{cetinkaya2025, cetinkaya2026}, which demonstrated the feasibility of intelligent mission coordination enabled by RL for a single mission with a smaller, homogeneous fleet and a fixed ignition point, this study extends the approach to multiple mission scenarios, a larger heterogeneous fleet, and varying ignition locations. We formulate mission execution as a Markov Decision Process and then apply Proximal Policy Optimization (PPO) - an RL-based algorithm \cite{schulman2017}- to train the coordinator. To illustrate the utility of the proposed approach, we use a simulation of a notional aerial firefighting mission developed by the German Aerospace Center (DLR) \cite{cigal2025}, in which a set of aerial firefighting assets is deployed to contain a wildfire within a designated region, subject to stochastic threat characteristics and environmental conditions, and their tactics are managed under the assumption of perfect, real-time information. By interacting with the simulation over thousands of episodes, the coordinator learns to map mission states to tactical actions, guided by reward signals that reflect the achievement of mission objectives. We train coordinators under two settings: single-scenario training, in which each coordinator is exposed to one mission scenario, and multi-scenario training, which aims to mitigate the brittleness of coordinators trained in isolation. 

\indent Our findings are three-fold. First, once trained, the coordinators significantly outperform two baselines: random tactics and a fixed heuristic tactic assignment, in the scenarios they were trained on. Second, single scenario training can be brittle. The coordinator trained on the more forgiving scenario falls to the level of random tactics when transferred to a more demanding one, whereas the coordinator trained on the demanding scenario retains most of its advantage when transferred, even though the fleet, its capabilities, and the available tactics are held constant and only the mission scenario changes. Third, within the same training budget, the multi-scenario coordinator performs on par with each single-scenario coordinator in that coordinator's own scenario and significantly outperforms all other policies when evaluated across both scenarios. The computational burden is incurred offline during training; once trained, the coordinator assigns tactics from the current mission state without re-optimizing at each decision point. While this paper utilizes an aerial firefighting case study, the proposed analytical approach is mission-agnostic in terms of its structure: the digital model, simulation, Markov Decision Process (MDP) formulation, and training pipeline carry over to other operational settings, although the state, action, and reward definitions must be re-derived for each mission. Ultimately, this study documents how coupling a DE-enabled mission simulation with advanced analytical tools can advance SE and ME practice concurrently.

\section{Literature Review}
This section reviews the three bodies of work on which this research builds: SoS and Mission Engineering, analytical methods for SoS Engineering (SoSE) and ME, and learning-based methodologies for SoSE.

\subsection{SoS and Mission Engineering} 
A System of Systems is a collection of systems that retain their operational and managerial independence while coming together to fulfill specific purposes \cite{maier1998}. An SoS is further characterized by evolutionary development, geographical distribution, and significantly more complex emergent behaviors compared to constituent systems acting alone \cite{maier1998, gorod2008}. Although SoSE is rooted in traditional SE, the tools, methods, and processes of SE exhibit several deficiencies when applied to SoS and cannot simply be transferred \cite{jamshidi2008}. These deficiencies stem from the defining characteristics of SoS: independently managed constituents require federated rather than centralized management, legacy systems constrain the SoS architecture, continuous evolution prevents the architecture from being finalized, and the scale of emergent behaviors has few analogies in traditional SE \cite{jamshidi2008}. Therefore, SoSE shifts from optimizing monolithic systems to orchestrating the interactions of a heterogeneous set of constituent systems to achieve capabilities beyond their individual scopes \cite{keating6, keating7, nielsen2015}, while accounting for uncertainty \cite{mcmanus2005, hastings2004, raz2020}.

However, this orchestration does not end at design time. To realize SoS objectives in changing environments, diverse sets of assets must be selected, configured, and coordinated to serve an evolving set of missions. Although SoSE methods have matured, deployed SoS architectures remain brittle \cite{gralla_characterizing_2016}: they are costly to change once implemented, and their performance is susceptible to operational and contextual changes \cite{raz2024ME, sapol2022}. Architecting an SoS of heterogeneous constituent systems, whose roles and level of integration may change dynamically with mission conditions, is challenging \cite{mitre2024}. Mission Engineering (ME) is an interdisciplinary process that bridges the gap between a mission and the engineering activities that support it, spanning the analysis, design, and integration of operational needs and capabilities toward desired mission outcomes \cite{dod2020ME}. Although the Department of War (DoW) definition is recent, the concept dates back at least to a 1993 NASA report that described ME as ``an end-to-end, multi-mission development methodology that seeks to integrate the development processes between the space, ground, science, and operations segments of a mission'' \cite{NASAME}. In both formulations, ME ties engineering activities to the mission environment they serve, which makes it well suited to addressing the brittleness of SoS architectures across application areas.

Raz et al. \cite{raz2024ME} propose a three-stage foundational ME framework for flexibly integrating and adaptively configuring SoS across multiple, evolving missions, addressing the limitations of traditional, static SoS architectures. The framework bridges conceptual modeling and analytical evaluation, providing a mathematical basis for managing the combinatorially explosive design space of dynamic, multi-mission environments. The authors further identify three ME needs: the ability to (1) rigorously and analytically navigate the SoS design space, (2) adaptively allocate tasks and reconfigure the SoS, and (3) identify multiple SoS configurations that can realize the mission. Because deployed SoS architectures are costly to modify, much of the adaptation burden falls on how the constituent systems are coordinated, which raises two difficulties: the tension between efficiency and the preservation of each constituent system's autonomy \cite{chen2024sosconductor}, and the brittleness of fixed coordination mechanisms and static strategies when conditions change \cite{chen2026adaptiveinfo}. Addressing these needs and difficulties calls for analytical methods that extend SoSE and ME from design into operation, enabling the SoS to adapt as missions and conditions evolve.

\subsection{SoSE and ME Using Analytical Methods}

One line of work applies computational tools to SoS architecting. DeLaurentis et al. (\cite{sosworkbench}) synthesized a workbench that maps each of the ``archetypal questions'' recurring in SoS architecting and evolution to a formal analytical tool. Within this line of work, Uday and Marais~\cite{wb5-uday2014resilience} define resilience-based system importance measures, which guide the design of more resilient SoS by providing metrics for analytical optimization. Davendralingam and DeLaurentis \cite{davendralingam2015robust} formulate constituent system selection as a financial portfolio optimization problem solved with robust optimization and, in a notional naval warfare scenario, generate a Pareto-optimal ``performance efficiency frontier'' that maps SoS performance across levels of accepted developmental risk. Furthermore, Guariniello et al.~\cite{wb4-guariniello2017supporting} builds on parametric extensions of functional dependency network analysis to propagate operability and schedule effects through dependency networks. Other relevant work in this area combines approximate dynamic programming with a transfer-contract coordination mechanism to plan multi-stage architecture evolution among conflicting stakeholders ~\cite{wb2-fang2014dynamic, wb3-fang2018multistakeholder}. Similarly, the architecting of Earth-observing satellite programs were formulated as three coupled combinatorial optimization problems composed of instrument selection, instrument-to-spacecraft assignment, and mission scheduling \cite{selva2014}; computing the non-additive benefit of the SoS with a rule-based expert system rather than linear programming ~\cite{selva2013vassar}. Taken together, these approaches establish a rigorous analytical basis for design-time SoS decisions, spanning capability--risk trade-offs, dependency-driven fragility, and multi-stakeholder evolution. However, they treat operations statically, leaving the operational coordination of constituent systems unaddressed.

A second line of research follows a similar series of SoS questions, from describing an SoS architecture, to selecting its constituents, to analyzing and protecting the network they form. The earliest work in this thread argues that static DoDAF-style architecture products cannot support dynamic analysis and propose generating executable models directly from architectural data through metamodel-level mapping rules \cite{executable2013}. Ge et al., treat constituent selection as a negotiation among stakeholders with divergent preferences \cite{ge2014interactive}. Building on earlier work of~\cite{hanDDnetwork, davendralingam2015robust}, Li et al. \cite{li2018networkportfolio} formulate portfolio selection as a multi-objective optimization in which capabilities emerge from network structure. Others cast SoS scheme generation as a path-ranking problem on a task--capability--system supernetwork \cite{supernet2015}. Moving to the network those constituents form, Li et al. \cite{structural-robustness} introduce an operation-loop representation and a spectral robustness measure, and Jiang et al. \cite{sos-killweb} extend this representation to kill webs (which resemble the mission webs of \cite{raz2024ME}) and derive a resilience index from the degradation and recovery of combat effectiveness under attack, concluding that decentralization and functional redundancy improve resilience. Liu et al. \cite{liu2023RL} use Q-learning to sequence protective investments against continuous disturbances. Collectively, this body of work provides representations and resilience metrics for SoS facing uncertainty (\textit{what} to build, \textit{what} is critical, and \textit{how} to protect it). However, like the first line, it treats mission operations as fixed inputs rather than as decisions to be made as the mission unfolds; even where RL is used~\citep{liu2023RL}, it sequences protective investments made before an attack rather than coordinating constituent systems during the mission. 

Beyond these two lines, several standalone studies apply analytical methods to SoS and ME problems. Raz et al.~\cite{raz2020} use machine learning to predict performance, robustness, and resilience of a smart-grid SoS to quantify their uncertainty. Covello et al. study network evolution empirically in a game environment, using graph-theoretic analysis rather than learning or optimization ~\cite{covello2023}. More recently, Huang et al. \cite{huang2026graphrag} combine large language models with knowledge graphs through graph retrieval-augmented generation to manage SoS complexity in an urban air mobility case. Most closely related to our work, Dachowicz et al.~\cite{dachowicz_mission_2021} propose an AI-infused framework that uses a real-time strategy game as a testbed to predict mission outcomes, quantify their uncertainty, and explain the factors driving them, concluding that combining uncertainty quantification with explainable AI closes the mission design loop.

\subsection{Learning-Based Methodologies for SoSE}

Learning-based methods, and RL in particular, offer a complementary route to SoS problems that call for adaptive decisions during operation rather than static, design-time analysis. In RL, an agent learns a policy ---a mapping from states to actions--- through trial-and-error interaction with its environment rather than explicit modelling, seeking to maximize the cumulative reward it receives over time~\cite{sutton1999, kaelbling1996reinforcement}. Early RL methods scaled poorly beyond low-dimensional problems because of their memory and computational complexity issues. Deep RL (DRL) overcomes this by approximating policies and value functions with neural networks, extending RL to previously intractable problems~\cite{arulkumaran2017deep}, beginning with Deep Q-Networks \cite{mnih2015} for high-dimensional discrete control tasks. However, value-based methods are documented to struggle with convergence in stochastic environments compared to policy-gradient methodologies \cite{sutton_policy_nodate} such as Trust Region Policy Optimization (TRPO) \cite{TRPO}, which constrains each update to guarantee monotonic improvement; and PPO \cite{PPO}, which achieves similar stability with a simpler clipped objective. 

A recent line of work by Chen, Heydari, and colleagues~\cite{chen2022dynamic, chen2024sosconductor, chen2025adaptiveinterv, chen2026adaptiveinfo, chen2024resourcegovernance, chen2026arch} applies hierarchical, multi-agent RL to the \emph{governance} of complex systems and SoS. The shared premise is that SoS constituents are autonomous, so a central authority cannot dictate their actions, yet system-level outcomes depend on how they operate collectively; governance therefore ``steers'' the SoS rather than commanding it. The pattern is two-tiered: constituents retain their decision making ability while a manager configures the SoS. The manager learns to allocate resources \cite{chen2022dynamic}, to intervene in the network structure linking the constituents \cite{chen2025adaptiveinterv, chen2026arch}, or to modulate the information each constituent receives \cite{chen2026adaptiveinfo}. Resource allocation has also been decentralized across subsystem managers drawing on a common pool \cite{chen2024sosconductor}.
In some cases, constituents first learn basic skills that are then frozen, before the manager is trained \cite{chen2022dynamic, chen2024sosconductor}. A related approach has been carried to competitive platform markets \cite{cao2025competing}. This body of work establishes that learning-based managers can steer autonomous constituents toward system-level objectives, which is precisely the relationship between our mission coordinator and the aircraft it coordinates. However, these studies learn in abstract environments (e.g., grid worlds, social dilemmas, random networks), and they do not connect the governance problem to mission outcomes.

Closer to the mission setting, although not framed in ME terms, several studies apply RL to mission-related planning objectives such as cooperative task allocation, tactical maneuver optimization, and dynamic threat avoidance~\cite{RL4UAV, RL4spaceplanning}. For instance, Yue et al. use PPO to train cooperative fighter--jammer UAV teams to coordinate their headings, velocities, and jamming sequences in continuous, high-dimensional air defense environments \cite{RL4UAV}. On the other hand, a modular training and testing framework was integrated with the event-driven SpaceSim platform, propagating high-fidelity orbital dynamics, communications, and remote sensing models across multi-satellite tracking and collision-avoidance scenarios \cite{RL4spaceplanning} .

\subsection{Synthesis and the Research Gap}

In summary, design-time analytical methods for SoSE and ME are mature but treat operations statically; learning-based governance methods address adaptive coordination of autonomous constituents but in abstract environments; and RL-based mission planners use realistic simulations but neither frame their problems in ME terms nor examine how the scenarios used in training shape performance after deployment. This paper addresses that gap by integrating RL with a high-fidelity digital mission model to train intelligent coordinators, framing the result as an ME capability, and explicitly studying the brittleness of the trained coordinators across mission scenarios.

\section{Methodology}

\begin{figure}[H]
\centering
\includegraphics[width=1.0\linewidth]{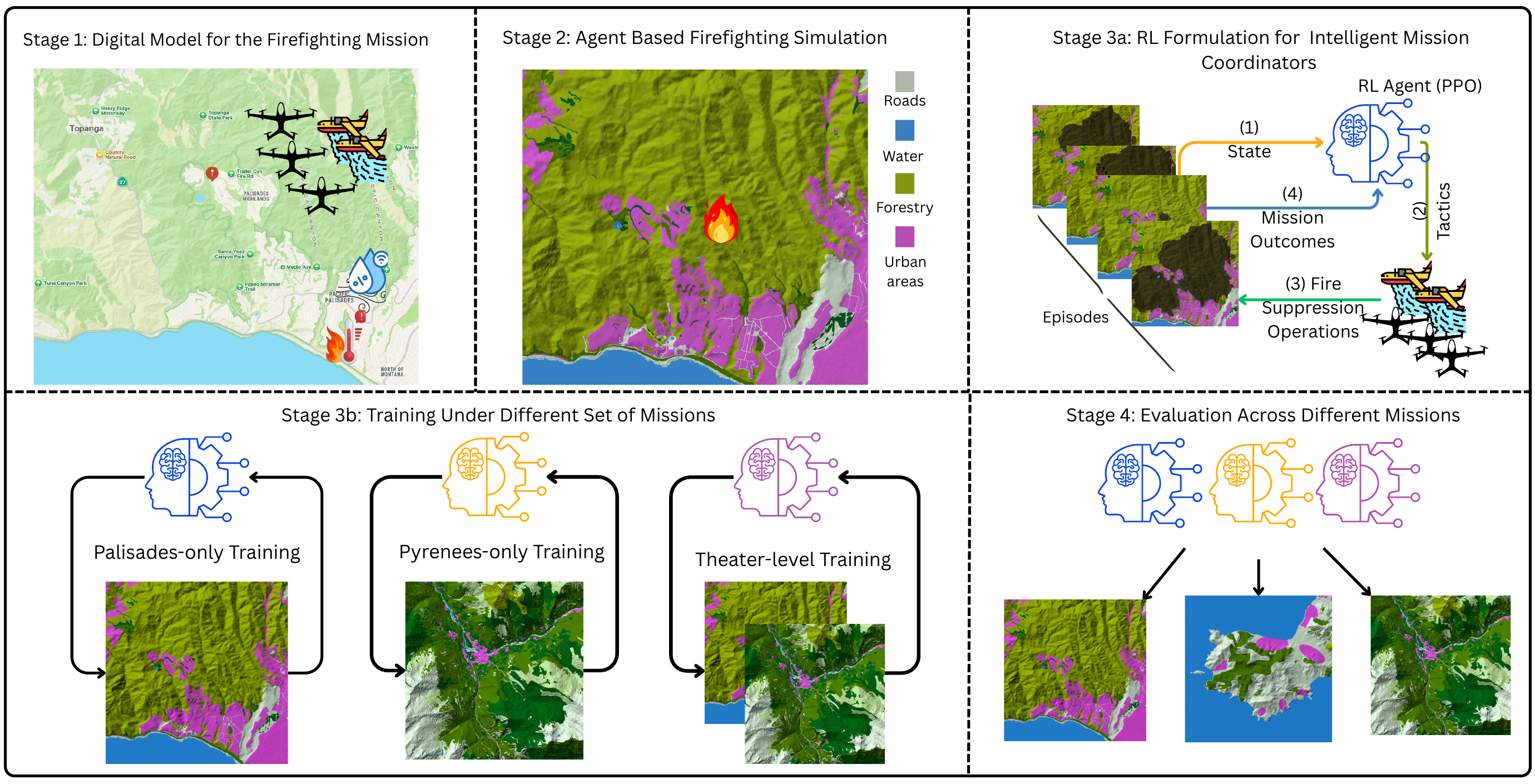}
\caption{Methodological Overview}
\label{fig:overview}
\end{figure}

We propose an intelligent mission coordination approach that enables real-time tactical decision-making by integrating high-fidelity digital mission models with RL. Figure \ref{fig:overview} provides an overview of our approach that consists of five stages. The first stage is a digital mission model that includes physics-based representations of the mission's geographical area, the threat, mission assets, laws of physics, and environmental conditions (e.g., vegetation type and density, wind, temperature) (Section~\ref{s:mission_model}). In the second stage, an agent-based simulation (ABS) built on this model, that links assets, environment, threat, and tactical decisions to simulate missions (Section~\ref{sec:ABS}). The third stage casts the mission coordination problem as an MDP and then builds on it to formulate RL-based intelligent mission coordinators for tactical decision-making (Section~\ref{s:MDP_RL}). In stage four, the coordinators are trained under two training settings: isolated scenario training and multi-scenario training (Section~\ref{s:training-workflow}). Finally we evaluate the agents through two different scenarios: Palisades and Pyrenees. We discuss each of these steps in detail throughout the remainder of this section.

\subsection{The Firefighting Mission Model}
\label{s:mission_model}

We utilize a digital mission model and an accompanying ABS developed by German Aerospace Center (DLR) as a testbed \cite{cigal2025}. The model comprises three elements: the geographical area, the firefighting fleet and its supporting infrastructure, and the mission scenarios that instantiate them.

\subsubsection{Geographical Area}
The model abstracts geographical areas by using a grid structure; where each cell is characterized with a set of attributes. Cells are of two types. \emph{Non-flammable} cells contain no fuel and therefore never ignite; they comprise roads, rocks, water sources, and airports, and are depicted in blue (water) and gray (all others) in Figures~\ref{fig:overview} and~\ref{fig:scenarios}. Because they never ignite, non-flammable cells act as natural firebreaks.\emph{Flammable} cells comprise urban areas (pink) and vegetation (green). Vegetation cells are further detailed by fuel type (e.g., pines, needles, fallen leaves, grasses, and pasture), so that flammability ranges from urban areas at the low end, through forest litter and pine stands, to grasses and pasture at the high end. These differences shape both fire propagation and mission outcomes: urban areas are harder to ignite than vegetation, but once they burn they incur heavy casualties and economic loss while contributing little to emissions. However, vegetation ignites and spreads fire easily and is the primary driver of emissions while contributing little to casualties and economic loss. At any time, each flammable cell is in one of five burning phases: combustible, early burning, full burning, extinguishing, or burnt.

\subsubsection{Firefighting Fleet and Infrastructure}
The fleet comprises two classes of aircraft, seaplanes and electric vertical take-off and landing aircraft (eVTOLs), whose characteristics are summarized in Table~\ref{tab:aircraft-comparison}. Both classes cruise at comparable speeds (100~m/s for seaplanes and 90~m/s for eVTOLs), so they differ mainly in payload, endurance, take-off \& landing  distances, and access to water. Each seaplane carries more than three times the payload of an eVTOL and can remain airborne far longer, but scooping requires a long, unobstructed stretch of open water, which limits seaplanes to large water bodies such as the ocean or large lakes. The eVTOLs have smaller payloads and shorter, battery-limited endurance, but because they take off and land vertically they can also reload from small water sources, such as ponds and reservoirs, that are inaccessible to seaplanes. These sources can lie closer to the fire, shortening the eVTOLs' turnaround between suppression runs. In this study the fleet is fixed at ten aircraft, five seaplanes and five eVTOLs, in every scenario. 

Airports are the only infrastructure elements in the mission model. They are represented by non-flammable cells and serve as the fleet's bases: aircraft take off from their assigned airport at deployment, return to an airport whenever their remaining fuel or battery charge requires refueling or recharging, and can also reload suppressant there as an alternative to water sources.

\begin{table}[H]
\centering
\caption{Characteristics of the seaplane and eVTOL platforms.}
\label{tab:aircraft-comparison}
\footnotesize
\setlength{\tabcolsep}{1pt}
\begin{tabular}{lcc@{\hspace{3em}}lcc}
\toprule
Parameter & Seaplane & eVTOL & Parameter & Seaplane & eVTOL \\
\cmidrule(r){1-3}\cmidrule(l){4-6}
Energy source         & Fuel            & Battery  & Cruise speed (m/s)       & 100     & 90   \\
Takeoff mode          & Runway or water & Vertical & Cruise altitude (m)      & 3{,}000 & 457  \\
Span (m)              & 18              & 6        & Takeoff distance (m)     & 2{,}760 & 0    \\
Payload capacity (kg) & 1{,}300         & 360      & Takeoff climb rate (m/s) & 3.81    & 0.51 \\
Endurance (min)       & 420             & 27       & Cruise climb rate (m/s)  & 6.96    & 3.56 \\
Range (km)            & 2{,}500         & 150      &                          &         &      \\
\bottomrule
\end{tabular}
\end{table}

\subsubsection{Mission Scenarios}
\label{s:scenarios}
\begin{figure}[H]
\centering
\includegraphics[width=0.7\linewidth]{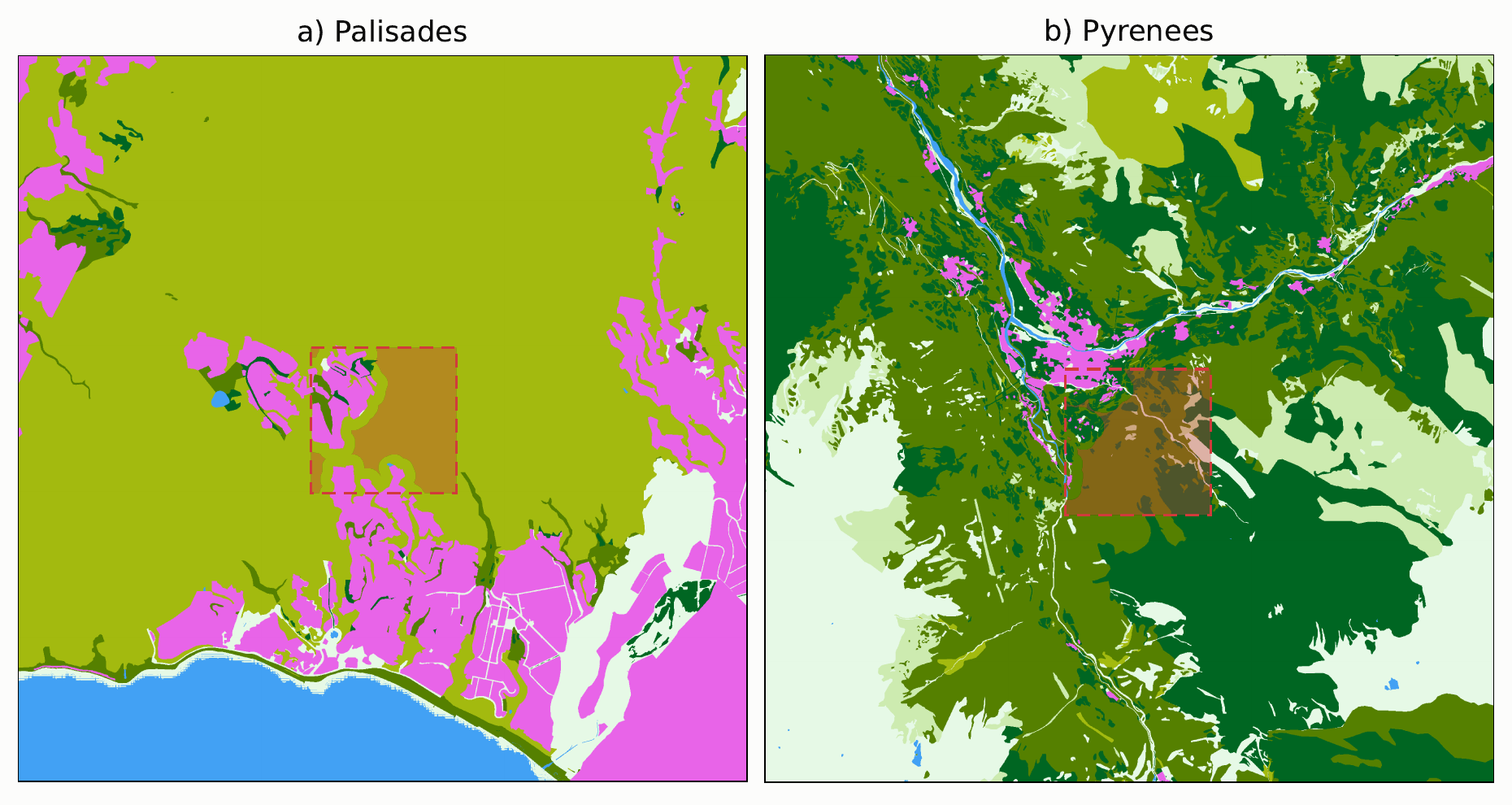}
\caption{Mission scenarios and the variable ignition regions for (a) Palisades and (b) Pyrenees.}
\label{fig:scenarios}
\end{figure}

This study adopts the mission model for two wildfire scenarios with distinct characteristics, both modeled by~\cite{cigal2025}: Palisades, a coastal California town, and Pyrenees, a small European town in a mountain range (Figure~\ref{fig:scenarios}). The Pyrenees scenario is hypothetical, whereas the Palisades scenario replicates the wildfire of January 7, 2025, a catastrophic event that burned 95~km$^2$, caused over 25 billion USD of damage, and resulted in 12 deaths over 31 days~\cite{babrauskas_palisades_2025}. Each scenario covers a $10 \times 10$~km area whose geography, characterized by proximity to water resources and a heterogeneous mix of urban areas and vegetation of varying density, presents a complex mission environment. In each episode, the fire ignites at a point drawn uniformly at random from a 4~km$^2$ ignition region within the scenario (Figure~\ref{fig:scenarios}); the proximity of the ignition point to urban or highly combustible areas affects both the difficulty of containment and the potential damage. The fire is detected after a scenario-specific response window, 1 hour in Palisades and 30 minutes in Pyrenees, after which the fleet is deployed.

Each scenario also specifies a nominal weather profile (Table ~\ref{tab:weather_ranges}). Temperature and relative humidity are defined as daily ranges and interpolated over time, peaking in the middle of the day. Wind speed fluctuates sinusoidally around its nominal value to represent natural variability, and wind direction is drawn from a uniform distribution around the scenario's prevailing direction. Wind speed and direction therefore vary from one time step to the next and constitute the main source of uncertainty in fire propagation, whereas temperature and humidity follow a more stable daily profile.

\begin{table}[htbp]
  \centering
  \caption{Nominal weather ranges of the Palisades and Pyrenees scenarios.}
  \label{tab:weather_ranges}
  \footnotesize
  \setlength{\tabcolsep}{1pt}
  \begin{tabular}{@{}lcc@{}}
    \toprule
    Parameter & Palisades & Pyrenees \\
    \midrule
    Temperature ($^{\circ}$C) & 10--23 & 10--25 \\
    Relative humidity (\%) & 22--90 & 20--70 \\
    Wind run (km/day) & 320 & 350 \\
    Wind speed (m/s) & 2.6--5.4 & 2.8--5.9 \\
    Wind direction ($^{\circ}$) & 15 $\pm$ 45 & 300 $\pm$ 30 \\
    \bottomrule
  \end{tabular}
\end{table}

\subsection{The Agent Based Simulation}
\label{sec:ABS}

The second stage is an ABS that links the behavior of the digital model elements over time. It integrates four elements: (i) the geographical area, (ii) the weather, (iii) the wildfire, and (iv) the mission assets; and simulates how they interact as a mission unfolds. The ABS determines when and where the fire ignites and how it propagates across the grid, and in parallel it simulates how each aircraft operates in this environment. Each water release alters how the fire spreads, so the fire and the fleet continuously shape each other, and mission outcomes emerge from the interaction of these elements rather than from any single one.

\subsubsection{Fire propagation}
The wildfire spreads from burning cells to neighboring flammable cells at a pace that depends on the cells' fuel type, the local terrain, and the current weather, advancing each cell through the five burning phases introduced in Section~\ref{s:mission_model}. Because of the variety in fuel, fire fronts spread unevenly, so the difficulty of containing a fire depends on the vegetation around it and on the evolving wind.

\subsubsection{Aircraft operations}
\label{sec:aircraft-ops}
To bound the scope of this study, firefighting is restricted to aerial intervention; no ground firefighting is modeled. Each aircraft is an independent agent with its own position, operational state, and class-specific performance parameters (Table~\ref{tab:aircraft-comparison}). At deployment, each aircraft departs its assigned airport fully loaded with suppressant and flies to the target determined by its current tactic, where it releases its payload, extinguishing the burning cells within its drop footprint. It then transits to a water source it can access (or an airport) to reload, flies to its next target, and releases again, repeating this cycle until its remaining fuel or battery charge requires a return to an airport for refueling or recharging, after which it resumes the cycle. Each aircraft engages a fire position determined by the tactic the coordinator assigns to it and by its class-level preferences. Given a point-of-interest (POI) selection tactic (Section~\ref{s:action}), each aircraft scores the ongoing fire positions against a set of operational priorities, such as its own proximity to the fire and the fire's proximity to protected objectives on the map, and engages the highest-scoring one. The priorities are combined through configurable weights that reflect each aircraft class's operational role.

\subsubsection{Mission Outcomes}
\label{sec:mission-outcomes}

At the end of each episode, four metrics are reported: (i) casualties (lives lost), (ii) cost of burnt area (\euro), (iii) burnt area (m$^2$), and (iv) emissions from the burnt area (tonnes CO$_2$). These are aggregated into a measure of effectiveness that serves as the coordinator's reward (Section~\ref{s:reward}).

\subsubsection{The Firefighting Fleet as an SoS}

Although tactics are assigned centrally, each aircraft selects its own targets: as described in Section \ref{sec:aircraft-ops}, each asset scores fire positions by combining operational priorities and class-specific weights and engages the highest-scoring one. These weights reflect the operational roles of seaplanes and eVTOLs and persist across scenarios, so the two classes retain distinct preference structures regardless of the coordinator's decisions.

The fleet thus satisfies the two principal criteria that Maier~\cite{maier1998} proposed for an SoS: operational and managerial independence. Each aircraft is operationally independent: it selects and suppresses targets without coordinating with other aircraft and would remain fully operable if the SoS were disassembled. The aircraft are also managerially independent, because their target selection follows class-specific preferences that the coordinator neither specifies nor overrides. These preferences reflect the diverse organizations from which constituent systems typically originate; for instance, the seaplanes might be operated by a federal agency such as the US Forest Service, whereas the eVTOLs might be owned by a regional authority such as the State of California.

Because the integrated fleet is assembled and centrally managed to fulfill a specific mission purpose, while its components subordinate their normal operational modes to that purpose only during mission execution, it is most accurately classified as a \emph{directed} SoS in Maier's taxonomy~\cite{maier1998}. The RL-based mission coordinator provides central management at the tactical layer by assigning each aircraft a tactic to coordinate the firefighting mission without revoking the components' operational or managerial independence: each aircraft still selects its own targets. Since this study aims to provide a proof of concept for the coordinator rather than to solve the fleet-selection problem, we deliberately deploy the same fixed fleet in every scenario and defer fleet planning to future work.

\subsection{Markov Decision Process Formulation and RL Implementation}
\label{s:MDP_RL}
We formulate the coordination problem as a MDP, a mathematical framework for modeling sequential decision making under uncertainty \cite{bellman1957markovian}, defined by the tuple $(\mathcal{S}, \mathcal{A}, P, \mathcal{R}, \gamma)$. The state space $\mathcal{S}$ represents all possible states of the world the agent may encounter; the action space $\mathcal{A}$ defines the alternatives available at each decision step; the transition probabilities $P$ describe how the environment evolves in response to the agent's actions; the reward function $\mathcal{R}$ provides a scalar signal of how desirable the outcome of each action is. Finally, the discount factor $\gamma \in [0, 1)$ determines how much the agent values future rewards relative to immediate ones. 

As the agent interacts with the environment, the system moves from one state to the next according to the transition probability $P(s_{t+1} \mid s_t, a_t)$ of reaching state $s_{t+1}$ after taking action $a_t$ in state $s_t$. The agent selects these actions through its policy $\pi(a_t \mid s_t)$, a distribution over actions conditioned on the current state. At each step, the agent receives a reward signal that is associated with the current state, the action taken, and the next state ($\mathcal{R}(s_t, a_t, s_{t+1})$); and it strives to maximize the expected long-term reward, discounted over a period ($\mathbb{E}_{\pi}[\sum_{t=0}^{\infty} \gamma^{t} * \mathcal{R}(s_t, a_t, s_{t+1})]$). In our implementation, the state captures the condition of the fire, the environment, and each aircraft; the actions are the tactics the coordinator assigns to the aircraft; and the reward is the change in a mission-level measure of effectiveness. Each component is defined below.

\subsubsection{The State Space}
\label{s:state}
To inform the coordinator about the current mission status, the state comprises 49 environment variables and 2 variables per aircraft at each time step (Table~\ref{tab:state-space} in the Appendix). The variables fall into five groups. The first identifies the mission scenario via a one-hot scenario identifier. The second describes mission progression through time since detection and time to sunset. The third captures the conditions that drive fire behavior: temperature, humidity, and wind speed and direction. The fourth provides a global description of the fire: its ignition point, centroid, and extreme points fix its location and extent; the dominant spread angle and the point at which the spread ray intersects the map boundary characterize its direction of propagation; and the shortest distances from the fire boundary to urban areas, water resources, highly combustible areas, high-slope terrain, and the indirect fire line capture its proximity to protected objectives. The fifth describes the fire at the sector level: the area around the fire center is partitioned into three directional sectors, and for each sector the state reports the severity of the fire front and five urgency scores quantifying the threat the sector poses to each class of protected objective. The final group records the current position of each of the ten aircraft. All variables are scaled to $[0,1]$ for training stability.

\subsubsection{The Action Space}
\label{s:action}
At each decision step the coordinator specifies a tactic for each aircraft. A tactic is a triplet:
\begin{itemize}
    \item \textbf{Select POI.} Points of interest are the target areas to be protected. This element dictates how the aircraft selects its POI from five alternatives: protect urban areas, protect water resources, protect highly combustible vegetation, protect areas where uphill spread is likely, or intervene indirectly by constructing a suppression line that contains the fire in an ellipse. Each alternative draws on its own subset of measures; for example, ``protect urban areas'' weighs the fire's proximity to urban zones against flight distance, whereas ``protect water resources'' additionally incorporates proximity to water sources. Aircraft executing the indirect alternative bypass this scoring and cooperatively construct the suppression line.
    \item \textbf{Track POI.} This element determines how the aircraft tracks its selected POI, with two alternatives: \emph{indirect} tracking, in which the aircraft recalculates a new POI once the initially selected point is completely burnt or suppressed, and \emph{follow the fire front}, which targets the point with the highest spread rate around the original POI. Following the fire front engages the fastest-spreading cells, but some suppressant may be dropped on areas that are not actively burning; it therefore trades payload utilization against target prioritization.
    \item \textbf{Suppress mode.} The final element defines how the suppressant is released: \emph{direct} suppression orients the drop to maximize the burning area captured while minimizing already-suppressed area, whereas \emph{indirect} suppression orients the drop tangentially to the fire's spread to complete the fire line.
\end{itemize}
These triplets yield 20 tactics per aircraft type (5 $\times$ 2 $\times$ 2). However, \cite{cigal2025} report that indirect tactics perform best in combination; accordingly, whenever the indirect option is chosen for Select POI, the indirect options for Track POI and Suppress mode are enforced. This collapses the four indirect combinations into a single valid tactic, leaving 17 tactics. Figure~\ref{fig:tactics} in Appendix \ref{sec:app-tactics} illustrates two of them.

\subsubsection{The Reward Structure Based on a Measure of Effectiveness}
\label{s:reward}
The reward is based on a notional measure of effectiveness (MoE) that captures the damage inflicted by the wildfire:
\begin{equation}
MoE_t = 1 - 0.25 \cdot \left( \frac{BA_t}{MBA_i} + \frac{CA_t}{MCA_i}
+ \frac{E_t}{ME_i} + \frac{C_t}{MC_i} \right) - \mathit{outofBound},
\label{eq:moe}
\end{equation}
where $BA_t$, $CA_t$, $E_t$, and $C_t$ are the burnt area, cost of the burnt area, emissions, and casualties accumulated up to time $t$, and $MBA_i$, $MCA_i$, $ME_i$, and $MC_i$ are the maximum damage each term can inflict in scenario $i$, bounded by the region's characteristics (e.g., maximum casualties equal the region's population; see Table~\ref{tab:moe-norms} in the Appendix). The binary term $\mathit{outofBound}$ equals 1 if the fire spreads beyond the region boundary and 0 otherwise. The MoE thus lies in $[-1, 1]$: values near 1 indicate successful missions with minimal losses, whereas negative values denote missions in which the fire escaped the region. The coordinator's reward at each step is the change in MoE, $r_t = \Delta MoE_t = MoE_t - MoE_{t-1}$, so that coordinator learns a policy maximizing the expected discounted sum of $\Delta MoE_t$ over an episode, $V^{\pi}(s_0) = \mathbb{E}_{\pi}[\sum_{t=1}^{T} \gamma^{t-1} \Delta MoE_t \mid s_0]$.

\subsubsection{Proximal Policy Optimization}
RL is a framework rather than a single method; once a problem is formulated as an MDP, algorithms ranging from value-based methods such as Q-learning~\cite{watkins1992q} and DQN~\cite{mnih2015} to policy-gradient methods such as REINFORCE~\cite{williams1992simple}, actor--critic~\cite{konda1999actor}, and TRPO~\cite{TRPO} can be applied. We employ PPO~\cite{PPO}, a policy-gradient algorithm valued for its training stability and ease of hyperparameter tuning, which handles discrete action spaces such as our tactic set effectively~\cite{tang2020discretizing}. PPO uses an actor--critic architecture: the actor $(\pi_{\theta}(a_t \mid s_t))$ is the policy from which actions are sampled, and the critic $(V_{\phi}(s_t))$ approximates the state-value function. PPO's stability stems from constraining each update to remain close, or proximal, to the previous policy. It computes the probability ratio $r_t(\theta) = \pi_{\theta}(a_t \mid s_t) / \pi_{\theta_{\mathrm{old}}}(a_t \mid s_t)$ and maximizes the clipped surrogate objective

\begin{equation}
L^{\mathrm{CLIP}}(\theta)
= \hat{\mathbb{E}}_{t}\!\left[
\min\!\Big(
r_t(\theta)\,\hat{A}_t,\;
\mathrm{clip}\!\big(r_t(\theta),\,1-\varepsilon,\,1+\varepsilon\big)\,\hat{A}_t
\Big)
\right],
\label{eq:eq_2}
\end{equation}
where $\hat{A}_t$ is the advantage estimate provided by the critic (how much better or worse the chosen action performed than the critic expected) and the clip operation prevents any single update from moving the policy far from $\pi_{\theta_{\mathrm{old}}}$. The advantage and critic losses follow the standard formulation and are given in Appendix~\ref{app:ppo}.

\subsection{Training of the RL Agent and Evaluation Design }
\label{s:training-workflow}

\subsubsection{Simulation Workflow}
\begin{figure}[htbp]
\centering
\includegraphics[width=1.0\linewidth, trim=40 125 40 125, clip]{"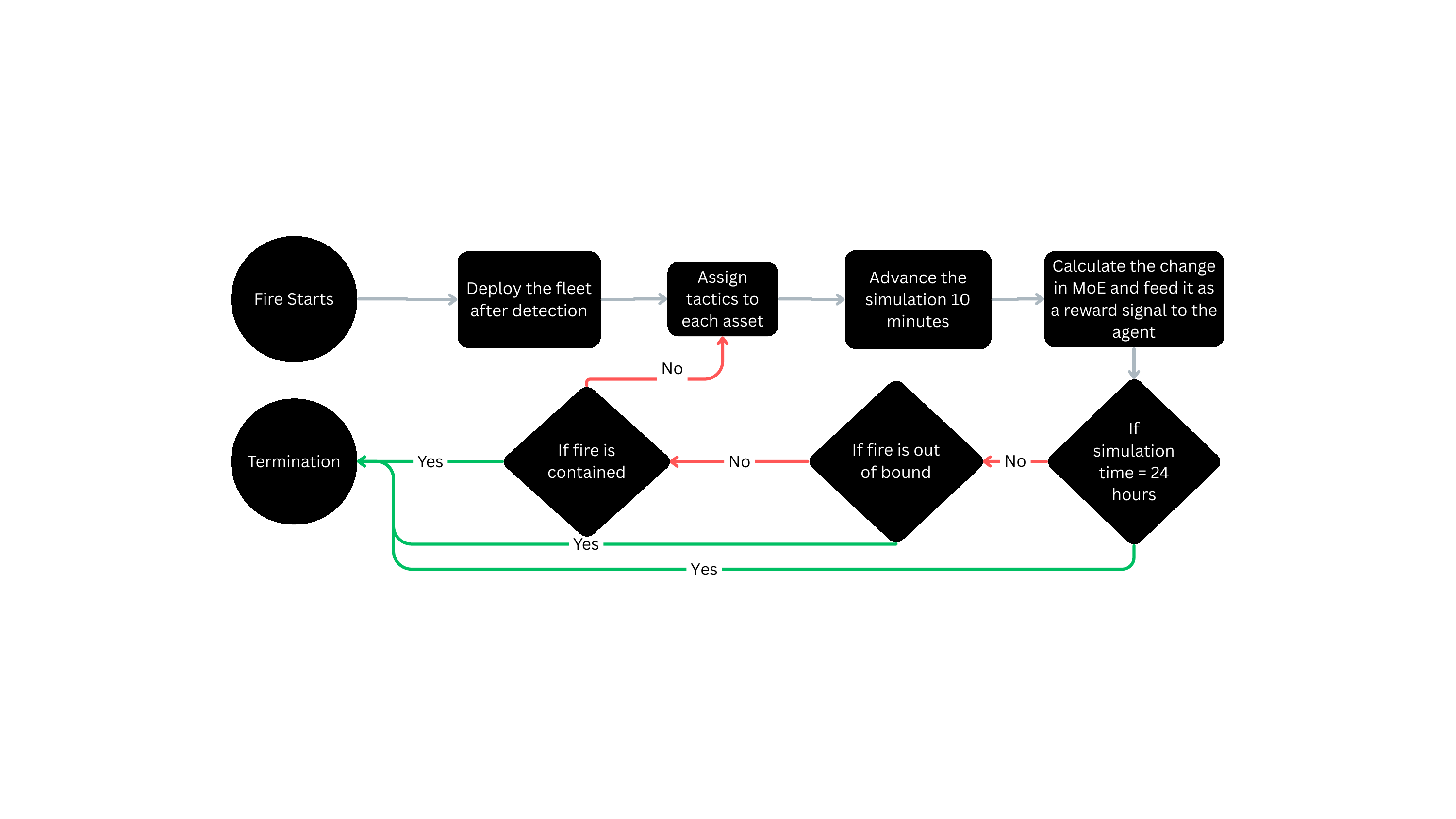"}
\caption{Operational workflow of the simulation environment for training the RL-based intelligent mission coordinator}
\label{fig:op_workflow}
\end{figure}

The simulation advances in discrete 10-minute time steps, which balance data collection against training time and are consistent with the cadence of aerial firefighting operations. Figure~\ref{fig:op_workflow} displays one episode: circles mark initiation and termination, boxes mark operations, and diamonds mark termination checks. The fire ignites, is detected after the scenario's response window, and the fleet is deployed. Every 10 minutes, the coordinator observes the state, assigns a tactic to each aircraft, and the simulation advances with those tactics; the resulting change in MoE is returned as the reward. The episode terminates when (i) the fire is contained, (ii) the 24-hour limit (144 decision steps) is reached, or (iii) the fire escapes the region boundary, meaning it will require further intervention by other resources; otherwise the cycle repeats.

\subsubsection{Training Setup}
\label{s:training_setup}
All coordinators were trained with PPO under an identical configuration. The actor and critic are separate networks sharing no weights, each with three hidden layers of 512, 256, and 128 units and tanh activations. Both were optimized with Adam, with the learning rate decaying linearly from $5 \times 10^{-4}$ to zero over 150{,}000 training episodes, a budget set by computational constraints. We set $\gamma = 0.99$, the GAE parameter $\lambda = 0.95$, the clipping range $\varepsilon = 0.2$, and the maximum gradient norm to 0.5. Training used 128 parallel workers on a single dual-socket AMD EPYC 9454 node with 96 physical cores and 755~GB of memory. Before each policy update, every worker contributes 144 decision steps, yielding a rollout of 18{,}432 samples that is shuffled into 24 mini-batches of 768 and optimized for 5 epochs (120 gradient steps per rollout); mini-batch updates reduce the variance of the gradient estimates and limit overfitting to specific simulation instances~\cite{schulman2017}.

Three coordinators are trained. The two single-scenario coordinators, Palisades-only (PA) and Pyrenees-only (PY), run all 128 workers on their respective scenario. The multi-scenario coordinator (MS) is trained on both scenarios concurrently. 64 workers are assigned to each scenario for the entire run, so every policy update is computed from a rollout containing an equal number of decision steps from each scenario. Because episodes that end early (contained or escaped fires) are shorter, the two scenarios contribute equal decision steps but unequal episode counts; over the full run, 44\% of MS's 150{,}000 episodes took place in Palisades and 56\% in Pyrenees. Alternative ways of dividing the budget are examined in Section~\ref{sec:curricula}.

\subsubsection{Baselines and Evaluation Protocol}
\label{s:eval_protocol}
Trained coordinators are evaluated with their policies held fixed on 1{,}000 episodes per scenario. The same 1{,}000 episodes (i.e., the same ignition points and weather realizations, generated from shared random seeds) are used for every policy, so that policies can be compared pairwise on identical fires. Two baselines are used. \emph{Random Tactics} assigns a new, randomly selected tactic to each aircraft every 10 minutes. \emph{Heuristics} is a fixed, role-based assignment held constant throughout the mission, directing the eVTOLs to protect water resources and the seaplanes to protect urban areas, which exploits each class's access to water and payload, respectively. This baseline represents the static, experience-based tasking rules on which emergency managers commonly rely~\cite{heuristics-paper, Fu2016, roberts2019}: context-sensitive guidelines that lead to satisfactory, though not necessarily optimal, solutions. Differences between policies are tested with two-sided paired t-tests on per-episode MoE, with Holm correction for multiple comparisons; we report the mean per-episode difference with its 95\% confidence interval.

\newpage 

\section{Findings}
In this section, we present our findings on the RL-based intelligent mission coordinators and is organized so that each subsection answers one research question. Section \ref{sec:rq1} addresses RQ1 by examining the training and evaluation performance of coordinators trained on a single scenario. Section \ref{sec:rq2} addresses RQ2 by evaluating each of these coordinators in a new scenario of the same mission they never encountered, and Section \ref{sec:rq3} addresses RQ3 by comparing multi-scenario training with single-scenario training. Finally, Section~\ref{sec:sensitivity-analysis} complements our findings with sensitivity analyses of ignition location, weather, and training curricula.

\subsection{RQ1: Integrating RL with a High-Fidelity Mission Model}
\label{sec:rq1}

\begin{figure}[H]
  \centering
  \includegraphics[width=0.55\linewidth]{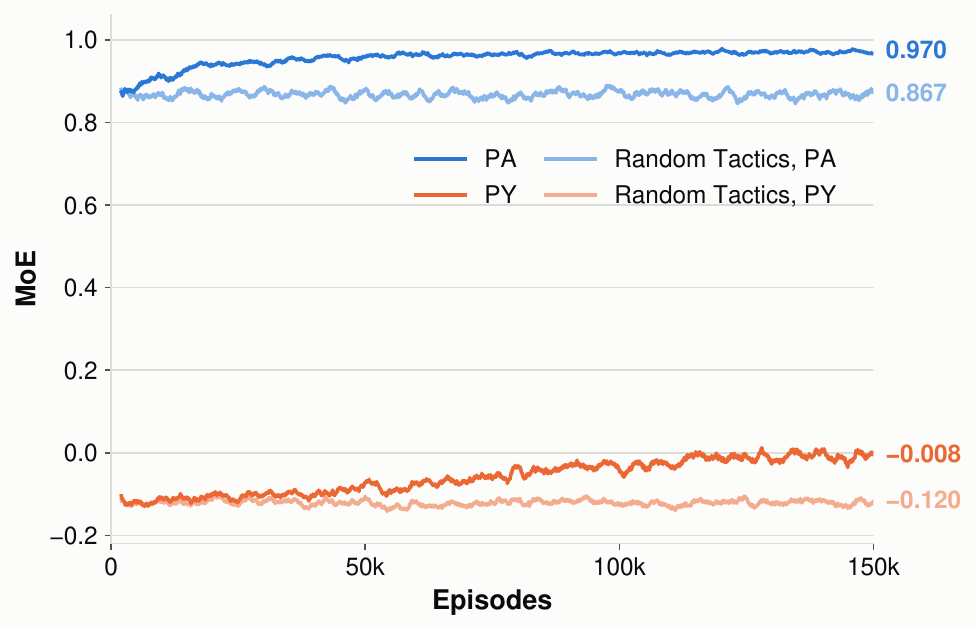}
  \caption{Training performance of PA and PY over 150{,}000 episodes}
  \label{fig:training-without-theater}
\end{figure}

To answer RQ1, we apply the proposed methodology, which integrates PPO with the high-fidelity DLR firefighting simulator, and examine the resulting intelligent mission coordinators. We first analyze how each coordinator learns when trained on a single scenario and then evaluate the trained coordinators against two baselines: random tactics and heuristics. Figure~\ref{fig:training-without-theater} presents results of two training runs under an identical budget of 150{,}000 episodes: PA, trained solely on Palisades (blue) and PY, trained solely on Pyrenees (orange). Bold lines show the PPO coordinators during training; pale lines of the same color show the Random Tactics baseline in the same scenario. The x-axis denotes the number of training episodes, the y-axis shows the 2{,}000-point moving average of the MoE, and the values at the right margin give each curve's average over last 20,000 episodes. 

Both coordinators outperform their random-tactics baselines, indicating that the coordinator successfully learns tactical policies that are superior to random assignment. Each PPO curve starts near its baseline and diverges upward as training progresses, while the baselines remain flat, as expected in the absence of learning. In Palisades, the coordinator improves rapidly, surpassing the baseline within the first 10{,}000 episodes and plateauing at approximately 0.97 after about 50{,}000 episodes while the baseline stays at 0.867. In Pyrenees, learning is slower and more gradual. The MoE rises steadily until roughly 110{,}000 episodes before leveling off near zero, finishing at $-0.008$, compared with $-0.120$ for the baseline. The absolute gains over the baselines are therefore similar in both scenarios (0.103 and 0.112), even though the two scenarios differ markedly in their achievable MoE levels, indicating that Pyrenees is a considerably more challenging mission environment. Negative MoE values in Pyrenees indicate that the fleet is undersized for this scenario, a property of the mission setup rather than of the coordination approach. In the context of our research questions, the relevant evidence of learning is not the absolute MoE level but the upward trend of the training curves and their separation from the random-tactics baseline. 

\subsection{RQ2: Brittleness of Single-Scenario Coordinators}
\label{sec:rq2}

\begin{figure}[H]
  \centering
  \begin{subfigure}[t]{\linewidth}
    \centering
    \includegraphics[width=0.65\linewidth]{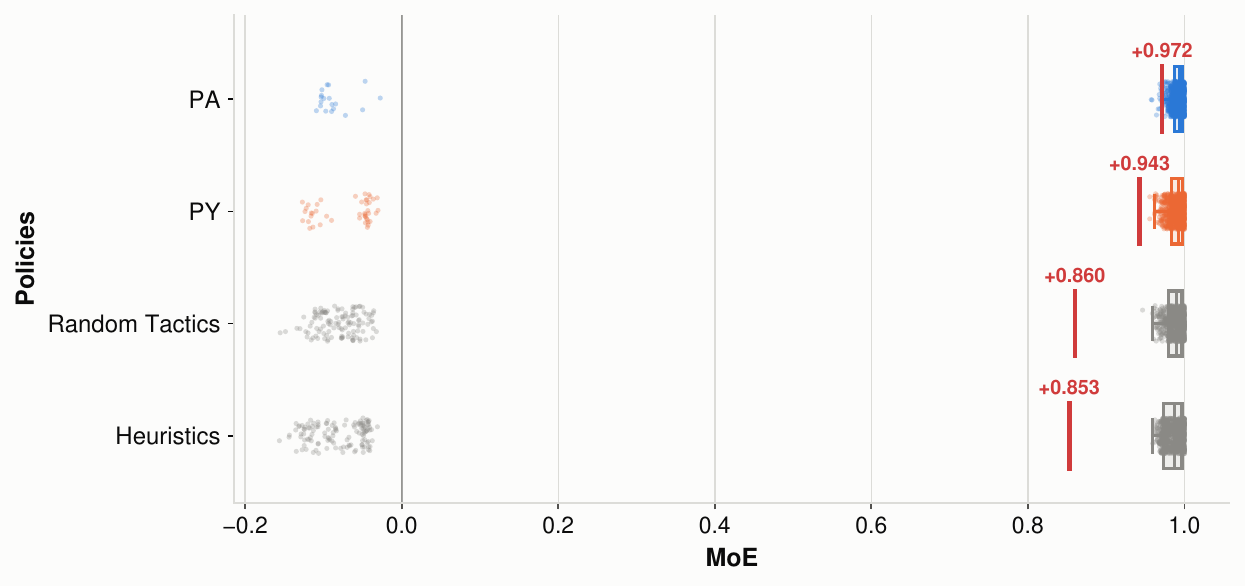}
    \caption{Palisades}
    \label{fig:eval_palisades_no_theater}
  \end{subfigure}

  \vspace{0.3em}
  \begin{subfigure}[t]{\linewidth}
    \centering
    \includegraphics[width=0.65\linewidth]{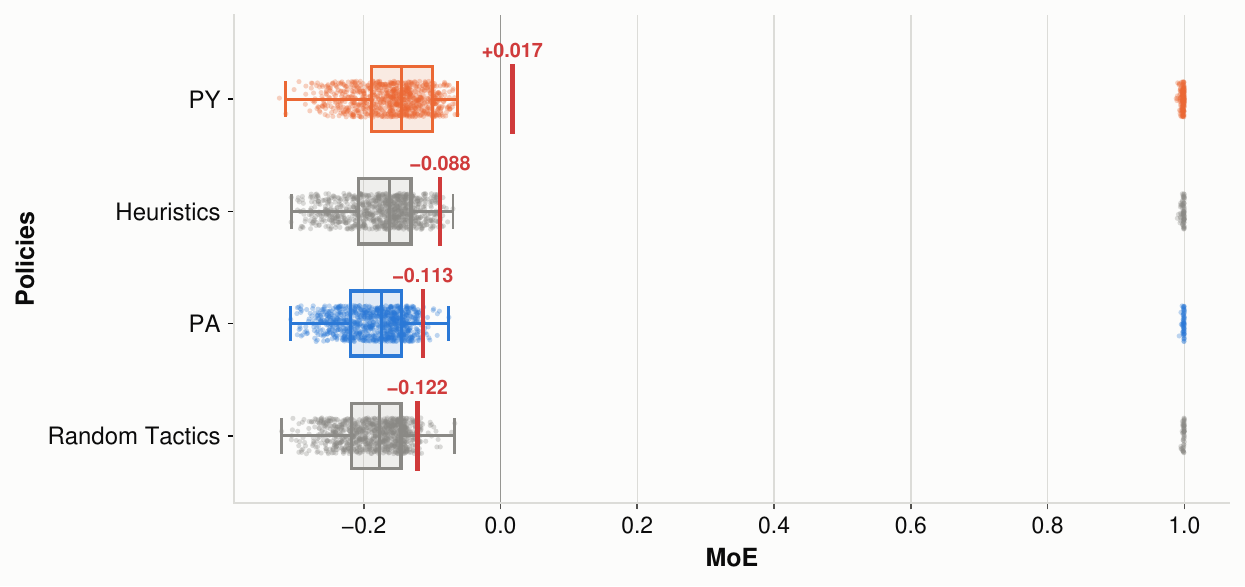}
    \caption{Pyrenees}
    \label{fig:eval_pyrenees_no_theater}
  \end{subfigure}

  \caption{MoE distributions of PY, PA and baselines across Palisades, and Pyrenees scenarios}
  \label{fig:eval_no_theater}
\end{figure}

To answer RQ2, each fully-trained single-scenario coordinator is evaluated in both scenarios, the one it was trained on and the one it never encountered, and compare their performance against both baselines. Figures ~\ref{fig:eval_palisades_no_theater} and \ref{fig:eval_pyrenees_no_theater} show the resulting MoE distributions, with red lines marking each policy's mean. The distributions are bimodal: missions either succeed, with MoE values close to 1, or fail, with negative MoE values when the fire escapes. Consequently, differences in mean MoE are driven largely by the proportion of contained fires rather than by variation within either mode (see Table \ref{tab:containment}). 

In Palisades (Figure~\ref{fig:eval_palisades_no_theater}), most missions succeed under all four policies, with MoE values clustered close to 1, while failed missions appear as outliers. PA achieves the highest mean MoE (0.972), followed by PY (0.943), both exceeding the random-tactics (0.860) and heuristics (0.853) baselines. While the difference between PA and PY is not statistically significant, both trained agents significantly outperform both baselines ($p_\mathrm{Holm} < 0.01$, see Table \ref{tab:stat-moe-comparisons}). Because the distributions of successful missions are closely similar across all four policies, coordinators' gains are rooted in a reduction of failures: the learned coordinators improve performance not by further enhancing already successful missions, but by avoiding the tactical choices that lead to mission failure (see Table \ref{tab:containment}). In Pyrenees (Figure~\ref{fig:eval_pyrenees_no_theater}), the pattern is reversed: most missions fail under every policy, reflecting the difficulty of this scenario. PY is the only policy that substantially outperforms the baselines, achieving a mean MoE of $0.017$, compared with $-0.122$ for random tactics and $-0.088$ for heuristics. PA, in contrast, performs no better ($-0.113$) than random tactics and heuristics ($-0.122$ versus $-0.088$, $p_\mathrm{Holm}>0.05$): its advantage in Palisades does not carry over to Pyrenees.

These results reveal an asymmetry in cross-scenario generalization. In Palisades, PY trails PA by 0.029 MoE; the difference between the two is not significant (0.943 versus 0.972), so PY retains most of its advantage in the scenario it never saw. In Pyrenees, by contrast, PA loses its advantage entirely, performing on par with Random Tactics. A plausible explanation lies in the difficulty of the two scenarios. Training in the demanding Pyrenees scenario forces PY to learn tactics that remain effective in easier settings, whereas Palisades is forgiving enough that even Random Tactics succeeds in most episodes, exerting limited pressure on PA to learn robust tactics. To answer RQ2, single-scenario training is brittle, and markedly so in one direction: the coordinator trained in in the forgiving scenario breaks down in the demanding one, whereas the coordinator trained in the demanding scenario degrades only slightly in the forgiving one.

\begin{table}[htbp]
\centering
\caption{Containment rate (share of the 1{,}000 evaluation episodes in which the fire did not escape the region) and mean MoE of each coordinator in each scenario}
\label{tab:containment}
\footnotesize
\setlength{\tabcolsep}{1pt}
\begin{tabular}{lcccc}
\toprule
 & \multicolumn{2}{c}{Palisades} & \multicolumn{2}{c}{Pyrenees} \\
\cmidrule(lr){2-3}\cmidrule(lr){4-5}
Policy & Contained (\%) & Mean MoE & Contained (\%) & Mean MoE \\
\midrule
PPO Palisades-only (PA) & 98.1 & 0.972 & 6.2 & $-0.113$ \\
PPO Pyrenees-only (PY)  & 95.5 & 0.943 & 15.5 & 0.017 \\
Heuristics              & 87.4 & 0.853 & 7.5 & $-0.088$ \\
Random Tactics          & 87.9 & 0.860 & 5.5 & $-0.122$ \\
\bottomrule
\end{tabular}
\end{table}

\subsection{RQ3: Multi-Scenario versus Single-Scenario Training}
\label{sec:rq3}

To answer RQ3, we compare the multi-scenario coordinator MS, trained jointly on Palisades and Pyrenees, under the same 150{,}000-episode budget (Section~\ref{s:training_setup}) and compare it with the single-scenario coordinators, PA and PY. In addition to the per-scenario comparison, we pool the two scenarios by combining each policy's 1{,}000 Palisades and 1{,}000 Pyrenees episodes into a single set of 2{,}000. Because both scenarios contribute equally, a policy's pooled mean MoE is the average of its two scenario means. The pooled view therefore measures how well a single policy performs across both scenarios, which is the relevant comparison for a coordinator that may be deployed in either one without knowing in advance which it will face. 

\subsubsection{Training}

\begin{figure}[H]
    \centering
    \includegraphics[width=0.60\linewidth]{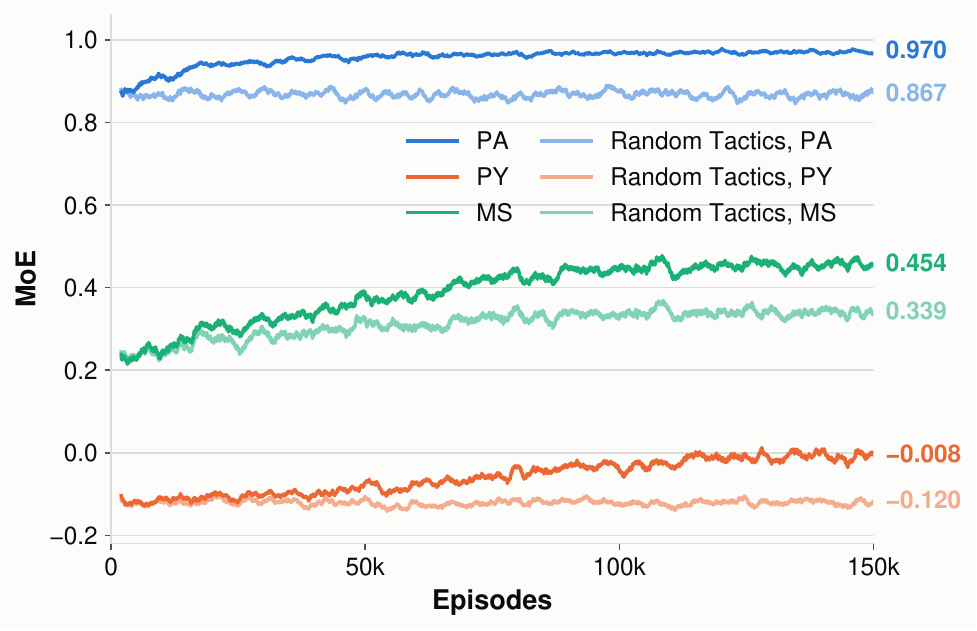}
    \caption{Comparative training performance of PA, PY, and MS over 150{,}000 episodes}
    \label{fig:train_all_instances}
\end{figure}

\begin{figure}[H]
    \centering
    \begin{subfigure}[b]{0.49\linewidth}
        \centering
        \includegraphics[width=\linewidth]{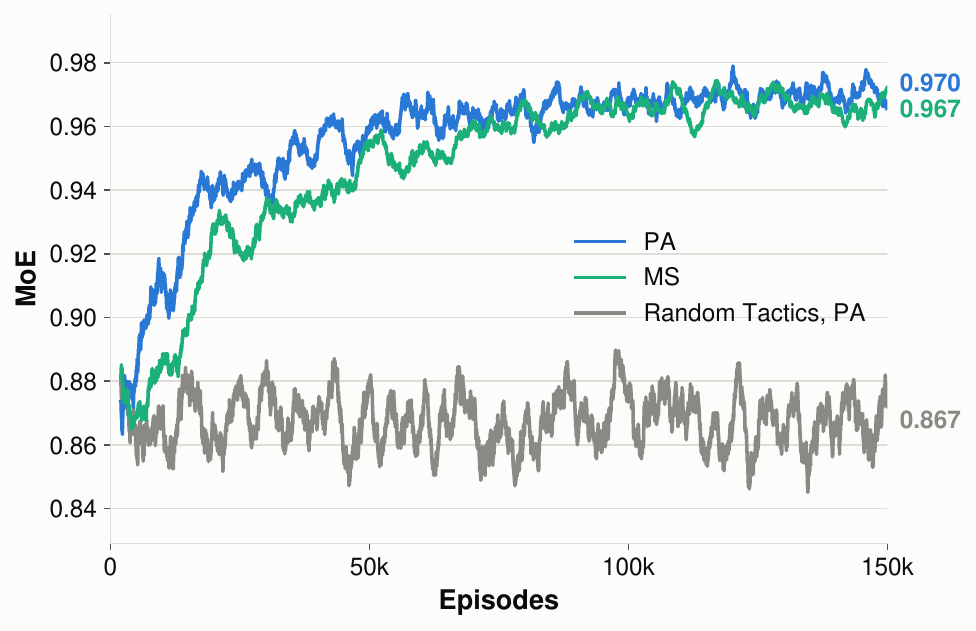}
        \caption{Palisades}
        \label{fig:train_ms_palisades}
    \end{subfigure}
    \hfill
    \begin{subfigure}[b]{0.49\linewidth}
        \centering
        \includegraphics[width=\linewidth]{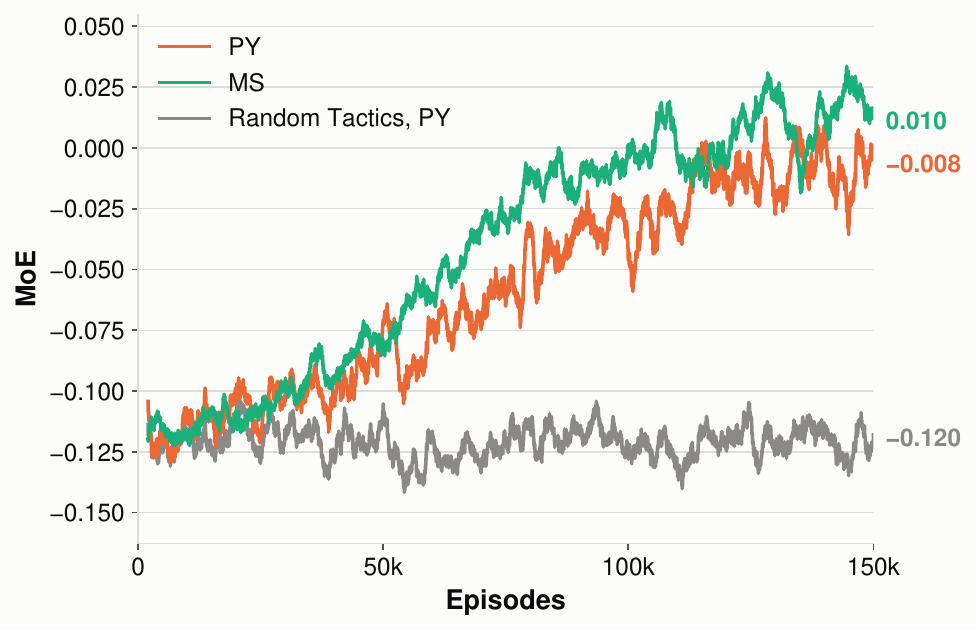}
        \caption{Pyrenees}
        \label{fig:train_ms_pyrenees}
    \end{subfigure}
    \caption{Comparative training performance of PA, PY, and MS over 150,000 episodes in (a) Palisades and (b) Pyrenees scenarios}
    \label{fig:train_ms_per_scenario}
\end{figure}

Figure~\ref{fig:train_all_instances} shows the training curves of all three coordinators with their corresponding random-tactics baselines. Because the MoE of MS is computed over episodes from both scenarios, its curve lies between those of PA and PY. MS concludes training with a mean MoE of 0.454, surpassing its random-tactics baseline (0.339). Although MS initially follows the baseline closely, it overtakes random tactics after approximately 30{,}000 episodes and continues its rise until 100{,}000 episodes, after which its progress slows. Figure~\ref{fig:train_ms_per_scenario} separates MS performance by scenario and compares it with the corresponding single-scenario coordinator. In these plots, each MS episode is placed at its position within the full concurrent run rather than at its count among that scenario's episodes, so both MS curves span the full 0--150{,}000 axis even though each covers fewer than 150{,}000 episodes. In Palisades (Figure~\ref{fig:train_ms_palisades}), MS converges to nearly the same MoE as PA (0.967 versus 0.970); the two follow different learning profiles initially but converge and track each other closely after 90{,}000 episodes.

 In Pyrenees (Figure~\ref{fig:train_ms_pyrenees}), MS matches and then exceeds PY during training despite running this scenario with only half of its workers: the two progress at a similar rate for roughly the first 50{,}000 episodes, after which MS rises above PY and remains above it for most of the remaining run, finishing at 0.010 versus $-0.008$. This hints at a potential benefit of multi-scenario training, namely that one coordinator can keep pace with the single-scenario coordinators in their own scenarios while dedicating only half of its training to each - a point we return to in Section \ref{sec:discussion}. Since training curves track an evolving policy through a moving average, we next compare the final coordinators under evaluation.

\subsubsection{Evaluation}
\begin{figure}[H]
    \centering
    \captionsetup[subfigure]{skip=1pt}
    \makebox[\linewidth][c]{%
    \begin{subfigure}[b]{0.57\linewidth}
        \centering
        \includegraphics[width=\linewidth]{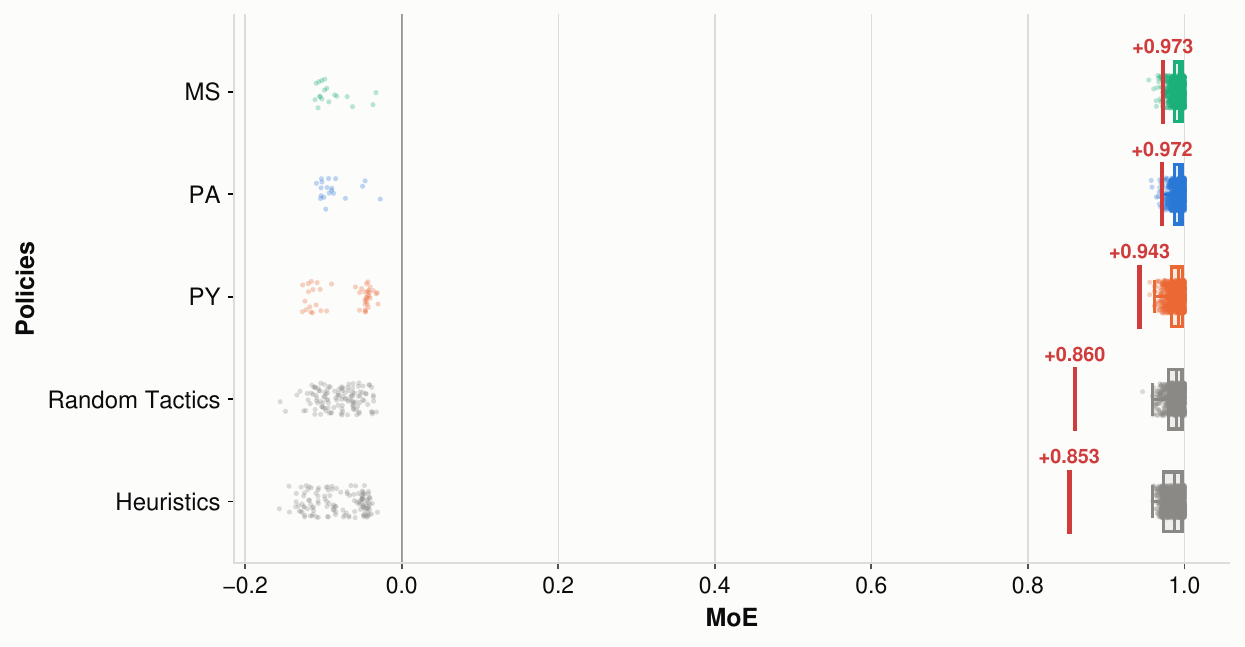}
        \caption{Palisades}
        \label{fig:eval_palisades_with_theater}
    \end{subfigure}\hspace{0.01\linewidth}%
    \begin{subfigure}[b]{0.57\linewidth}
        \centering
        \includegraphics[width=\linewidth]{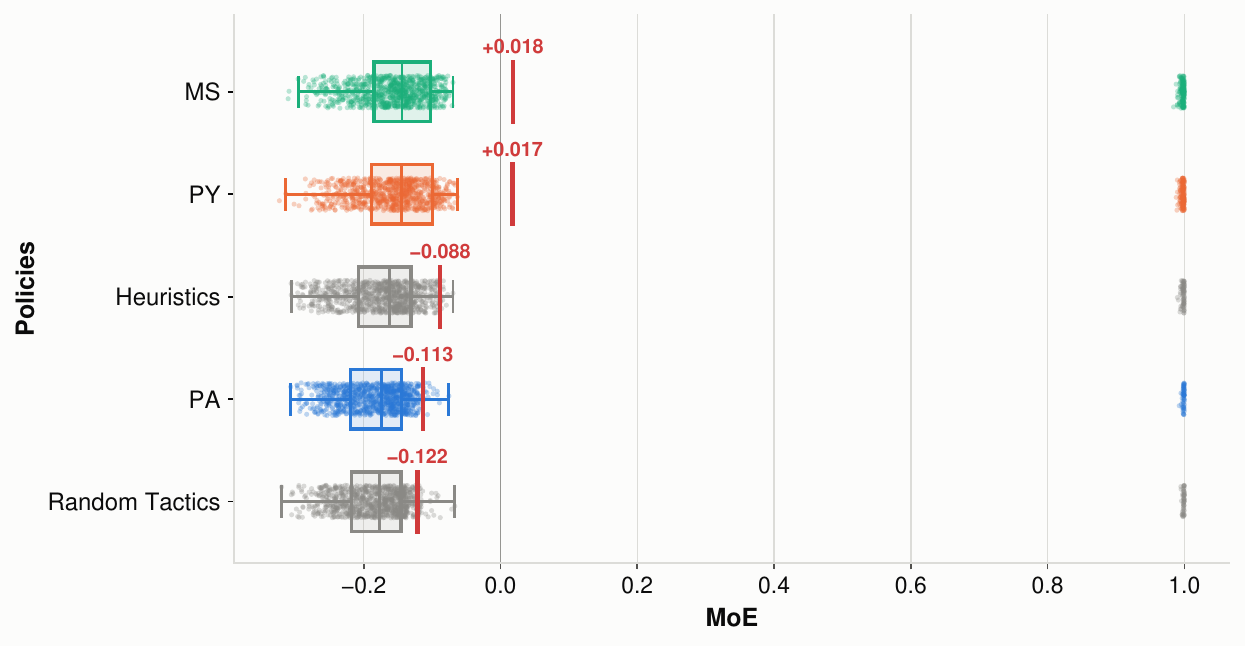}
        \caption{Pyrenees}
        \label{fig:eval_pyrenees_with_theater}
    \end{subfigure}}

    \vspace{2pt}
    \begin{subfigure}[b]{0.57\linewidth}
        \centering
        \includegraphics[width=\linewidth]{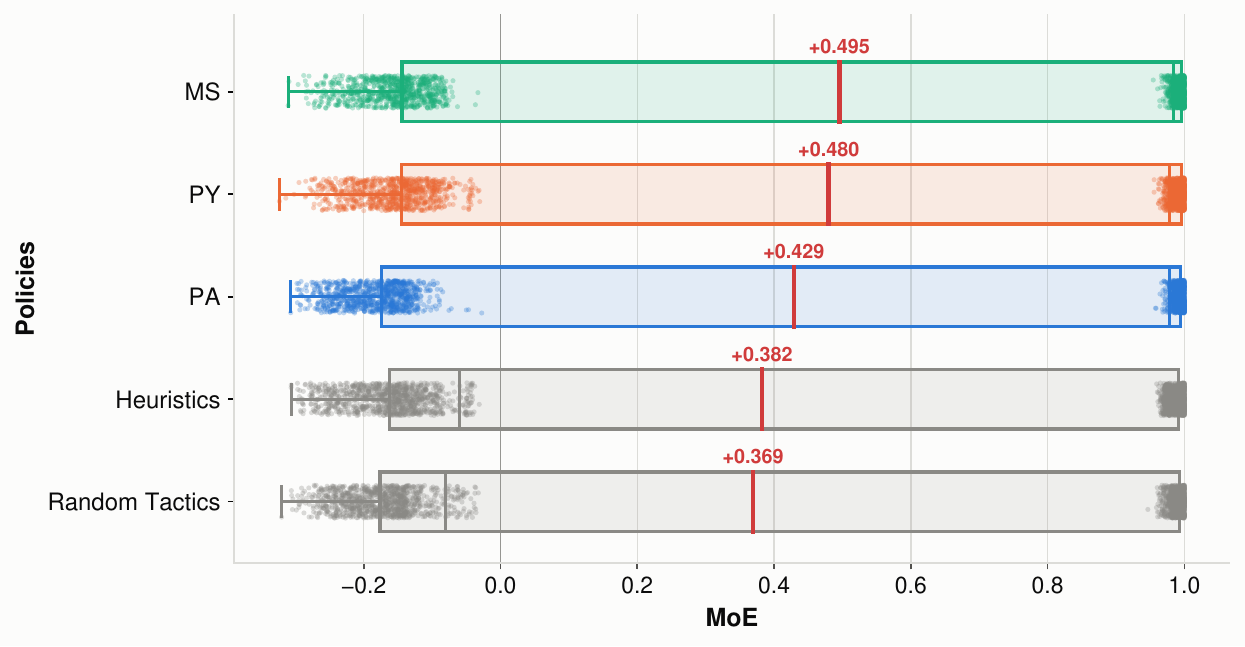}
        \caption{Pooled}
        \label{fig:eval_pooled_with_theater}
    \end{subfigure}
    \caption{MoE distribution of MS, PY, PA, and baselines across (a) Palisades, (b) Pyrenees scenarios, and (c) Pooled across both over 1{,}000 evaluation shared episodes on each scenario}
    \label{fig:eval_with_theater}
\end{figure}

Figure~\ref{fig:eval_with_theater} extends the evaluation of Section~\ref{sec:rq2} to include MS, comparing the MoE distributions of PA, PY, MS, Heuristics and Random Tactics over the same 1,000 simulated fires (episodes) per scenario, with red lines marking each policy's mean MoE.

MS achieves the highest mean MoE in each scenario as well as pooled across both. Within each scenario, MS is statistically indistinguishable from the coordinator trained on that scenario, whereas pooled across both scenarios it significantly outperforms every other policy (Table~\ref{tab:stat-moe-comparisons}). In Palisades (Figure~\ref{fig:eval_palisades_with_theater}), MS attains a mean MoE of 0.973, almost identical to PA (0.972, $p_{\mathrm{Holm}}=1$). Both significantly outperform PY (0.943, $p_{\mathrm{Holm}}=0.0012$), Random Tactics, and Heuristics (0.860 and 0.853, $p_{\mathrm{Holm}}<0.0001$)). In Pyrenees (Figure~\ref{fig:eval_pyrenees_with_theater}), MS attains the highest mean MoE (0.018), followed by PY (0.017), with no significant difference between them ($p_{\mathrm{Holm}}=0.92$); MS's advantage over PY during training thus does not translate into a significant difference in evaluation. Both significantly outperform PA ($-0.113$), which falls below Heuristics ($-0.088$) and is statistically indistinguishable from Random Tactics ($-0.122$). Pooled across both scenarios (Figure~\ref{fig:eval_pooled_with_theater}), MS achieves the highest mean MoE (0.495), significantly ahead of PY (0.480; $p_{\mathrm{Holm}}=0.033$), PA (0.429), Heuristics (0.382), and Random Tactics (0.369), the latter three with $p_{\mathrm{Holm}} < 0.001$. Although MS's per-scenario lead over PY is not significant in Pyrenees, MS leads PY by 0.030 in Palisades and 0.001 in Pyrenees, and these leads combine into a significant pooled advantage. PA, by contrast, falls behind because of its poor performance in Pyrenees.

\begin{table}[H]
\centering
\caption{Pairwise statistical comparisons of coordinators' MoE in Palisades, Pyrenees, and both scenarios pooled using paired $t$-tests with Holm correction}
\label{tab:stat-moe-comparisons}
\footnotesize
\begin{tabular*}{\textwidth}{@{\extracolsep{\fill}}lrcr@{}}
\toprule
A vs B & $\Delta$MoE & 95\% CI & $p_\mathrm{Holm}$ \\
\midrule
\multicolumn{4}{@{}l}{\textit{Palisades ($1{,}000$ simulations)}} \\
\quad MS vs PA & ${+}0.0010$ & [$-0.0099$, ${+}0.0120$] & $1$ \\
\quad MS vs PY & ${+}0.0299$ & [${+}0.0137$, ${+}0.0460$] & $0.0012$ \\
\quad PA vs PY & ${+}0.0288$ & [${+}0.0131$, ${+}0.0446$] & $0.0012$ \\
\quad PY vs Heuristics & ${+}0.0899$ & [${+}0.0684$, ${+}0.1113$] & $3.6\times10^{-15}$ \\
\quad PY vs Random Tactics & ${+}0.0826$ & [${+}0.0593$, ${+}0.1058$] & $2.8\times10^{-11}$ \\
\midrule
\multicolumn{4}{@{}l}{\textit{Pyrenees ($1{,}000$ simulations)}} \\
\quad MS vs PY & ${+}0.0009$ & [$-0.0184$, ${+}0.0202$] & $0.92$ \\
\quad MS vs PA & ${+}0.1317$ & [${+}0.1088$, ${+}0.1545$] & $4.6\times10^{-27}$ \\
\quad PY vs PA & ${+}0.1307$ & [${+}0.1085$, ${+}0.1530$] & $4.5\times10^{-28}$ \\
\quad Heuristics vs PA & ${+}0.0253$ & [${+}0.0090$, ${+}0.0416$] & $0.0072$ \\
\quad PA vs Random Tactics & ${+}0.0081$ & [$-0.0052$, ${+}0.0215$] & $0.46$ \\
\midrule
\multicolumn{4}{@{}l}{\textit{Pooled ($2{,}000$ simulations)}} \\
\quad MS vs PY & ${+}0.0154$ & [${+}0.0028$, ${+}0.0280$] & $0.033$ \\
\quad MS vs PA & ${+}0.0663$ & [${+}0.0534$, ${+}0.0793$] & $2.7\times10^{-22}$ \\
\quad PY vs PA & ${+}0.0509$ & [${+}0.0369$, ${+}0.0650$] & $6.8\times10^{-12}$ \\
\quad PA vs Heuristics & ${+}0.0467$ & [${+}0.0325$, ${+}0.0609$] & $3.9\times10^{-10}$ \\
\quad PA vs Random Tactics & ${+}0.0598$ & [${+}0.0464$, ${+}0.0731$] & $1.7\times10^{-17}$ \\
\bottomrule
\end{tabular*}
\end{table}

\subsubsection{Multi Scenario Training Leads to Lowest Overall Regret}

To compare the approaches within and across scenarios, we compute the regret of each approach's mean outcome in three scenarios: Palisades, Pyrenees, and both scenarios pooled, where all 2{,}000 episodes carry equal weight. Let $\bar{m}_c(a)$ denote the mean MoE of approach $a$ over the episodes in scenario $c \in \{\mathrm{Palisades}, \mathrm{Pyrenees}, \mathrm{pooled}\}$, and let $\mathcal{A}$ be the set of five evaluated approaches. The regret of approach $a$ in scenario $c$ is:
\begin{equation}
R_c(a) = \max_{a' \in \mathcal{A}} \bar{m}_c(a') - \bar{m}_c(a),
\label{eq:regret}
\end{equation}
that is, its shortfall from the best of the five approaches in that scenario, so the best approach has zero regret and lower regret is better. We refer to the regret in the pooled as the overall regret.

\begin{figure}[H]
    \centering
    \includegraphics[width=1.0\linewidth]{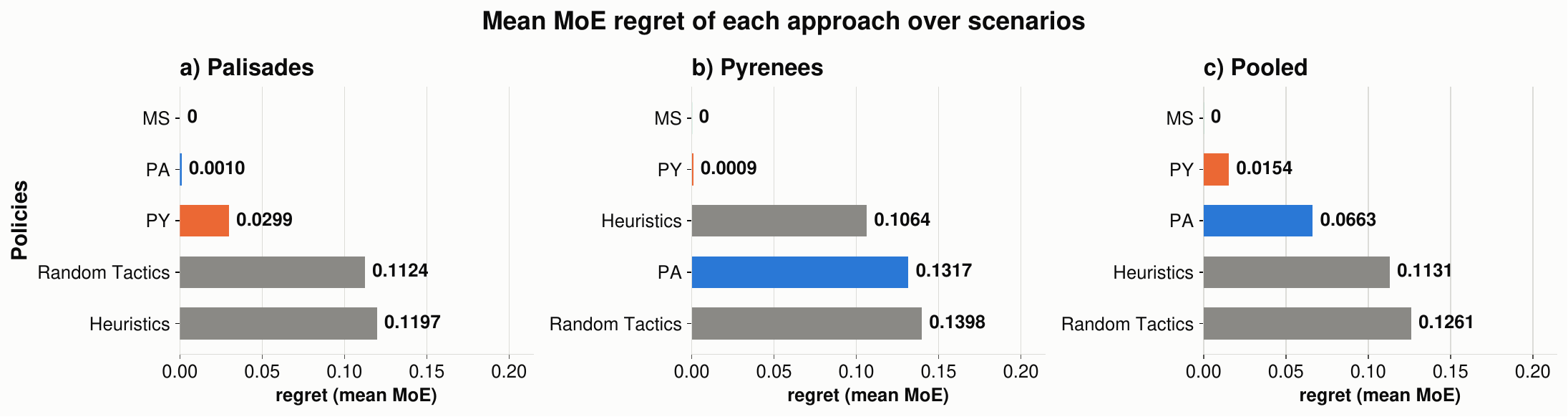}
    \caption{MoE regret of coordinators in (a) Palisades, (b) Pyrenees, and (c) pooled across both scenarios (overall regret).}
    \label{fig:regret}
\end{figure}

Figure~\ref{fig:regret} illustrates that MS has zero regret in all three contexts. In Palisades (Figure~\ref{fig:regret}a), MS attains the highest mean MoE and therefore zero regret, closely followed by PA (0.0010) and PY (0.0299), whereas the and Random Tactics and Heuristics baselines trail by 0.1124 and 0.1197, respectively. And in Pyrenees (Figure~\ref{fig:regret}b), MS again has zero regret, and PY trails it by 0.0009. PA, by contrast, has larger regret (0.1317) compared to Heuristics (0.1064) and a lower regret compared to Random Tactics (0.1398). Overall, (Figure~\ref{fig:regret}c), MS attains zero regret, followed by PY (0.0154), PA (0.0663), and the Heuristics (0.1131) and random-tactics (0.1261) baselines. 

Taken together, the paired statistical tests and the regret analysis show that, within the same simulation budget, the coordinator trained jointly on both scenarios performs on par with the native single-scenario coordinator in each scenario, attains the highest mean MoE in every context and, significantly outperforms every other policy when pooled. This is notable given that Palisades experience alone did not transfer to Pyrenees. The specialization of PA to Palisades leaves it with the largest coordinator regret in Pyrenees, whereas MS avoids this brittleness. To answer RQ3, multi-scenario training recovers the brittleness of single-scenario coordinators across the scenarios included in training without sacrificing per-scenario performance. Whether it also improves performance in scenarios absent from training is not tested here and is addressed in Section~\ref{sec:discussion}.

\subsection{Sensitivity Analysis}
\label{sec:sensitivity-analysis}

\subsubsection{Sensitivity to the Ignition Location During Training}
\label{sec:sensitivity}

All coordinators in the preceding analyses were trained on fires whose ignition points vary within a 4~km$^2$ region. To examine how sensitive the learned coordinators are to this choice, we retrained PY with a fixed ignition point at the center of that region and denote the resulting variant PY$_{\mathrm{fix}}$; the ignition rule is the only difference from PY (Section~\ref{s:training_setup}).

\begin{figure}[H]
    \centering
    \begin{subfigure}[b]{0.50\linewidth}
        \centering
        \includegraphics[width=\linewidth]{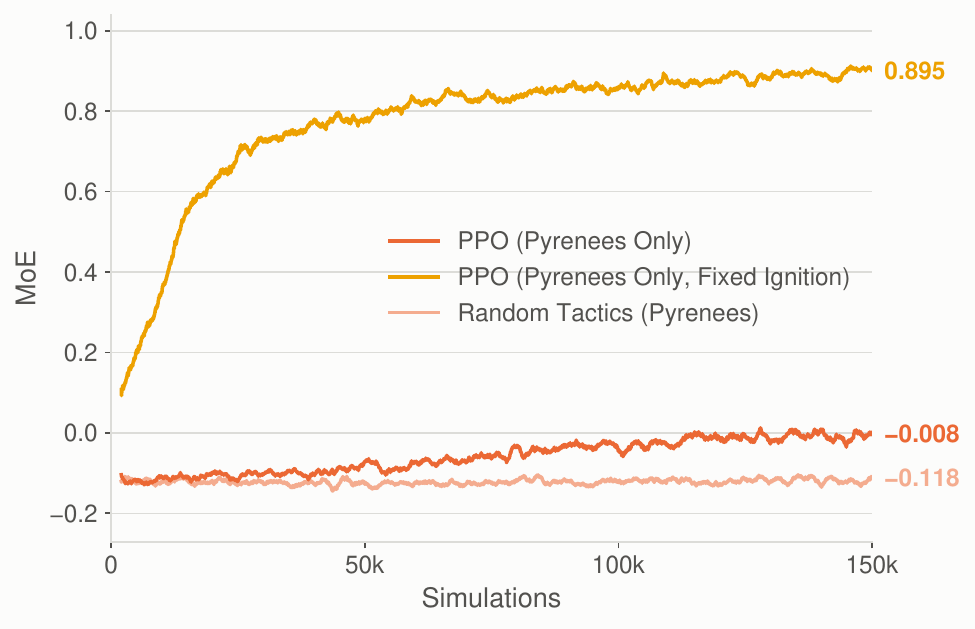}
        \caption{Training in Pyrenees}
        \label{fig:sens_train_pyrenees}
    \end{subfigure}%
    \hfill
    \begin{subfigure}[b]{0.50\linewidth}
        \centering
        \includegraphics[width=\linewidth]{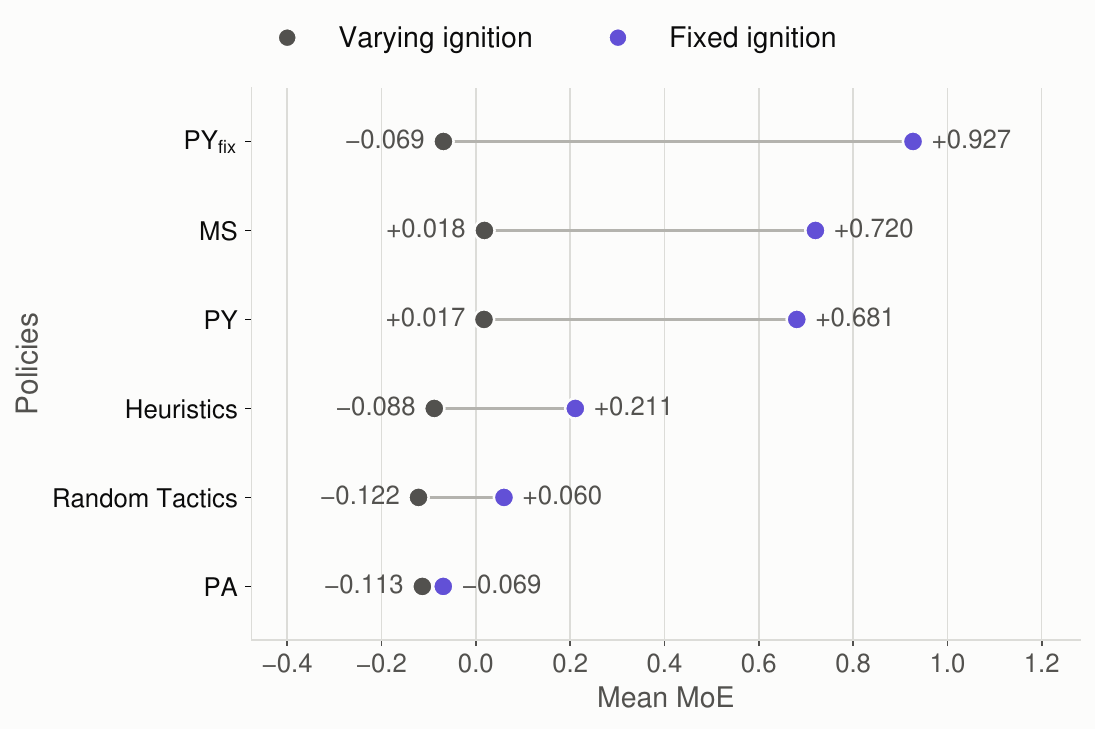}
        \caption{Evaluation in Pyrenees}
        \label{fig:sens_eval_pyrenees}
    \end{subfigure}
    \caption{(a)~Training MoE of PY and PY$_{\mathrm{fix}}$ over 150{,}000 simulations, with the Random Tactics baseline for reference. (b)~Mean evaluation MoE of PA, PY, MS, PY$_{\mathrm{fix}}$, and the baselines under varying and fixed ignition (each dumbbell connects a policy's scores in the two settings).}
    \label{fig:sens_eval}
\end{figure}

Figure~\ref{fig:sens_train_pyrenees} compares the training curves. PY$_{\mathrm{fix}}$ rises steeply during roughly the first 25{,}000 episodes and finishes training at an MoE of 0.895, far above both PY ($-0.008$) and the Random Tactics baseline. These curves measure performance on the fires encountered during training, so the MoE of PY$_{\mathrm{fix}}$ reflects a single, repeated fire, while PY reflects tens of thousands of different ignition points. Evaluation under varying ignition points reveals the limits of this advantage (Figure~\ref{fig:sens_eval_pyrenees}): PY$_{\mathrm{fix}}$'s mean MoE falls from 0.927 under fixed ignition to $-0.069$ under varying ignition Although it still significantly outperforms PA ($-0.113$), Heuristics ($-0.088$) and Random Tactics ($-0.122$), it falls significantly short of PY ($0.017$) and MS ($0.018$). Both perform well without ever having been trained on that specific fire, with MS again ahead.

The contrast between the training and evaluation performance of PY$_{\mathrm{fix}}$ indicates that a coordinator trained on a single ignition point learns tactics tailored to that ignition point rather than to the scenario as a whole. From an ME perspective, this result underscores that the threat characteristics used to train a coordinator, such as the ignition location, should reflect the variability expected in operation. Otherwise, a coordinator that appears highly effective during training may overfit to static conditions and prove brittle once deployed.

\subsubsection{Sensitivity to Adverse Weather Conditions}
\label{sec:sens_weather}

To examine how the coordinators respond to changes in environmental conditions, we evaluate them without retraining, under adverse weather and mild weather variants of each scenario obtained by modifying three weather parameters relative to the nominal weather (Table~\ref{tab:weather_ranges}). In the adverse variant, the daily temperature curve is shifted up by 5\,$^{\circ}$C, the relative humidity range is lowered by 10 percentage points at both ends, and the daily wind run is increased by 20\% in adverse weather. Whereas in the mild weather regime, the daily temperature curve is shifted down by 5,$^{\circ}$C, the relative humidity range is raised by 5 percentage points at both ends, and the daily wind run is reduced by 10\%. Ignition points, the fleet, water sources, wind direction, and all timing parameters are unchanged.

\begin{figure}[tb]
    \centering
    \begin{subfigure}[b]{0.50\linewidth}
        \centering
        \includegraphics[width=\linewidth]{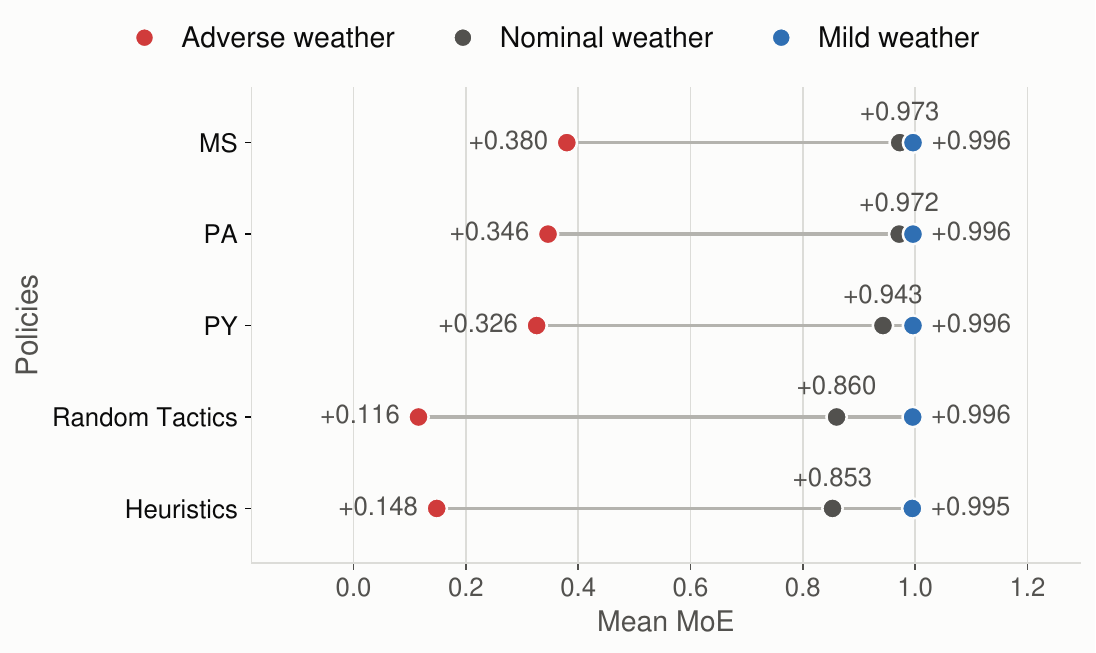}
        \caption{Palisades}
        \label{fig:sens_weather_palisades}
    \end{subfigure}%
    \hfill
    \begin{subfigure}[b]{0.50\linewidth}
        \centering
        \includegraphics[width=\linewidth]{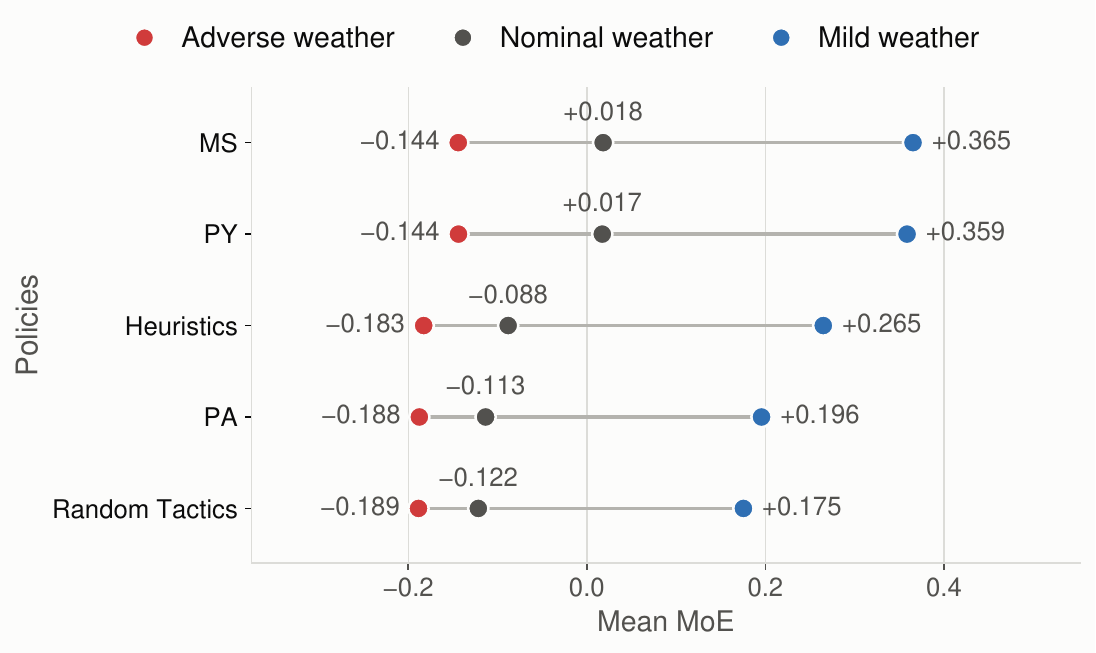}
        \caption{Pyrenees}
        \label{fig:sens_weather_pyrenees}
    \end{subfigure}
    \caption{Mean evaluation MoE of coordinators and the baselines under adverse and mild weather conditions (each dumbbell connects a coordinator's scores in the three settings).}
    \label{fig:sens_weather}
\end{figure}

Weather conditions shift the performance of every policy but largely preserve their relative standing (Figure~\ref{fig:sens_weather}). Adverse weather lowers the mean MoE of every policy in both scenarios, mild weather raises it, and MS matches or exceeds every other policy under all three conditions. In Palisades (Figure~\ref{fig:sens_weather_palisades}), adverse weather causes the steepest decline: the mean MoE falls from 0.943--0.973 to 0.326--0.380 for the coordinators and from 0.853--0.860 to 0.116--0.148 for the baselines. MS attains the highest mean MoE (0.380), followed by PA (0.346) and PY (0.326), while Heuristics and Random Tactics reach only 0.148 and 0.116; the margin between the coordinators and the baselines therefore widens, with MS leading the stronger baseline by 0.232 compared with 0.113 under nominal weather. Mild weather has the opposite effect, lifting every policy, including the baselines, to 0.995--0.996, so coordination makes virtually no difference when conditions are this favorable. In Pyrenees (Figure~\ref{fig:sens_weather_pyrenees}), where most fires already escape under nominal weather, adverse weather pushes every mean MoE further below zero. MS and PY retain the highest means (both $-0.144$), ahead of Heuristics ($-0.183$), PA ($-0.188$), and Random Tactics ($-0.189$), but their margin over the baselines narrows to 0.039 from 0.106 under nominal weather. Mild weather lifts every policy above zero while preserving the nominal ranking, with MS (0.365) and PY (0.359) ahead of Heuristics (0.265), PA (0.196), and Random Tactics (0.175).

Taken together, the relative standing of the policies largely persists across weather conditions, but the value added by the learned coordinators depends on those conditions: it vanishes when conditions are favorable enough for any policy to succeed (Palisades under mild weather), shrinks when conditions overwhelm the fleet (Pyrenees under adverse weather), and is largest in between, most notably in Palisades under adverse weather. From an ME perspective, SoS performance thus depends on operational conditions as well as on the coordination approach: because adverse weather was not seen during training, a harsh change in conditions after deployment sharply degrades the absolute performance of even well-trained coordinators. Their relative advantage over the baselines, however, carries over to conditions not seen during training.

\subsubsection{Sensitivity to the Multi-Scenario Training Curriculum}
\label{sec:curricula}

A further factor that can influence the RQ3 findings is how the multi-scenario training budget is divided between scenarios. In the main results, MS was trained concurrently, with half of its parallel workers assigned to each scenario; and because simulation times differ between the scenarios, this yielded 44\% of the training simulations in Palisades and 56\% in Pyrenees. Because the scenarios differ markedly in difficulty, an even allocation is not necessarily the best choice. To test the sensitivity of our findings to this choice, we examined two sequential alternatives under the same 150{,}000 episode budget. The first keeps the even balance, with 75{,}000 simulations in Palisades followed by 75{,}000 in Pyrenees; the second devotes most of the budget to the more demanding scenario, with a warm-up of 25{,}000 episodes in Palisades followed by 125{,}000 episodes in Pyrenees. Each resulting coordinator is evaluated on the same 1{,}000 episodes per scenario, (Table~\ref{tab:eval-curricula}).

Concurrent training outperforms both alternatives, attaining the best mean MoE in both scenarios and pooled, whereas the even sequential split performs worst overall; both sequential curricula fall below the single-scenario coordinators in their own scenarios. Because the even sequential split divides the budget about as evenly as concurrent training does, its poor performance indicates that the order in which the scenarios are presented matters, not only the share of the budget each receives. It is plausible that training on the scenarios one after the other might be leading the coordinator to partially forget the first. Multi-scenario training is therefore not sufficient on its own; it must be paired with an appropriate allocation and scheduling of the training budget to be effective.

\begin{table}[H]
\centering
\caption{Mean MoE of multi-scenario coordinators trained under three curricula that divide the same 150{,}000-simulation budget differently between Palisades and Pyrenees.}
\label{tab:eval-curricula}
\footnotesize
\setlength{\tabcolsep}{1pt}
\begin{tabular}{lrrr}
\toprule
 & \multicolumn{3}{c}{Mean MoE} \\
\cmidrule(l){2-4}
Curriculum & Palisades & Pyrenees & Pooled \\
\midrule
Sequential, even (75k $\rightarrow$ 75k)     & 0.9409 & $-$0.0062 & 0.4673 \\
Sequential, warm-up (25k $\rightarrow$ 125k) & 0.9480 & 0.0150 & 0.4815 \\
Concurrent (44\% / 56\%)                     & \textbf{0.9726} & \textbf{0.0182} & \textbf{0.4954} \\
\bottomrule
\end{tabular}
\end{table}

\subsubsection{Sensitivity to an Unseen Scenario}
\label{sec:unseen-scenario}

\begin{figure}[H]
    \centering
    \includegraphics[width=0.35\linewidth]{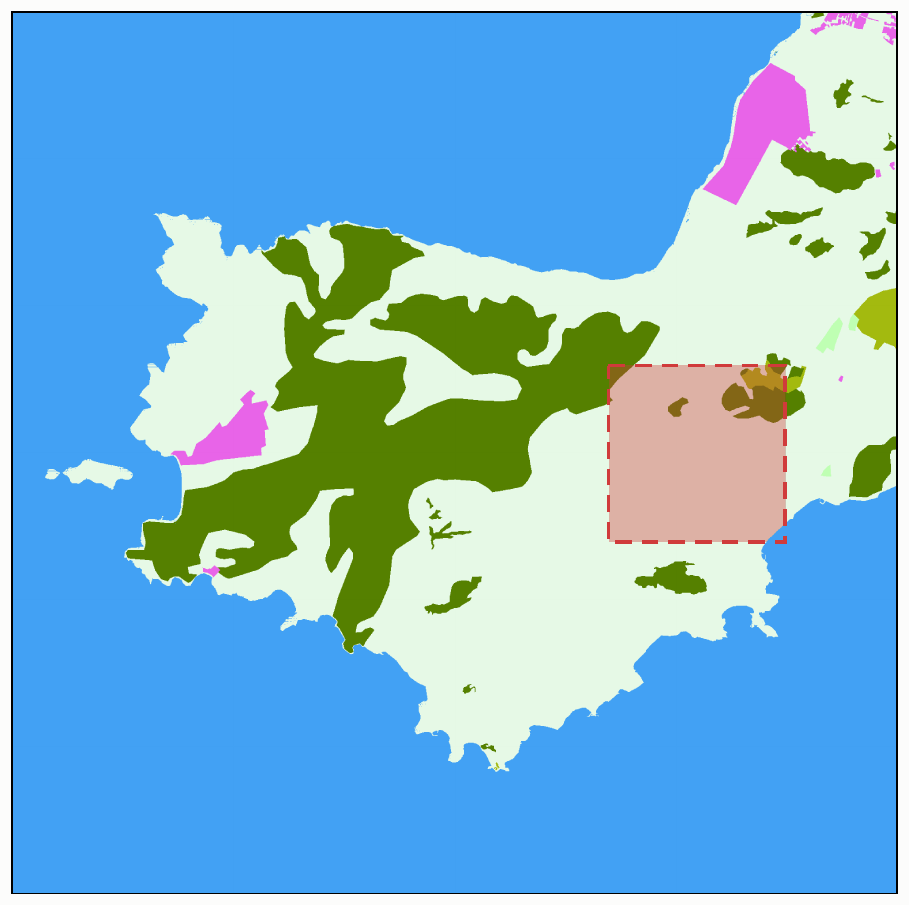}
    \caption{Salamis scenario}
    \label{fig:salamis-map}
\end{figure}

One of the core arguments of this article is the brittleness of ML algorithms and limits to their generalizability. To investigate how the coordinators respond to a new scenario that was not in the training set, we evaluated them in a third scenario: Salamis, an Aegean island (Figure \ref{fig:salamis-map}. As in the training scenarios, each simulation draws its ignition point from a 4~km$^2$ region, and every coordinator is evaluated on the same 1{,}000 simulations. Salamis proves more forgiving than Pyrenees but harder than Palisades: every coordinator attains a higher mean MoE than in Pyrenees and a lower one compared to Palisades. Its intermediate difficulty also makes the choice of policy matter more: mean MoE spans 0.381 to 0.824 across coordinators in Salamis, compared with 0.853 to 0.973 in Palisades and $-0.122$ to $0.018$ in Pyrenees. 

\begin{figure}[H]
    \centering
    \includegraphics[width=0.65\linewidth]{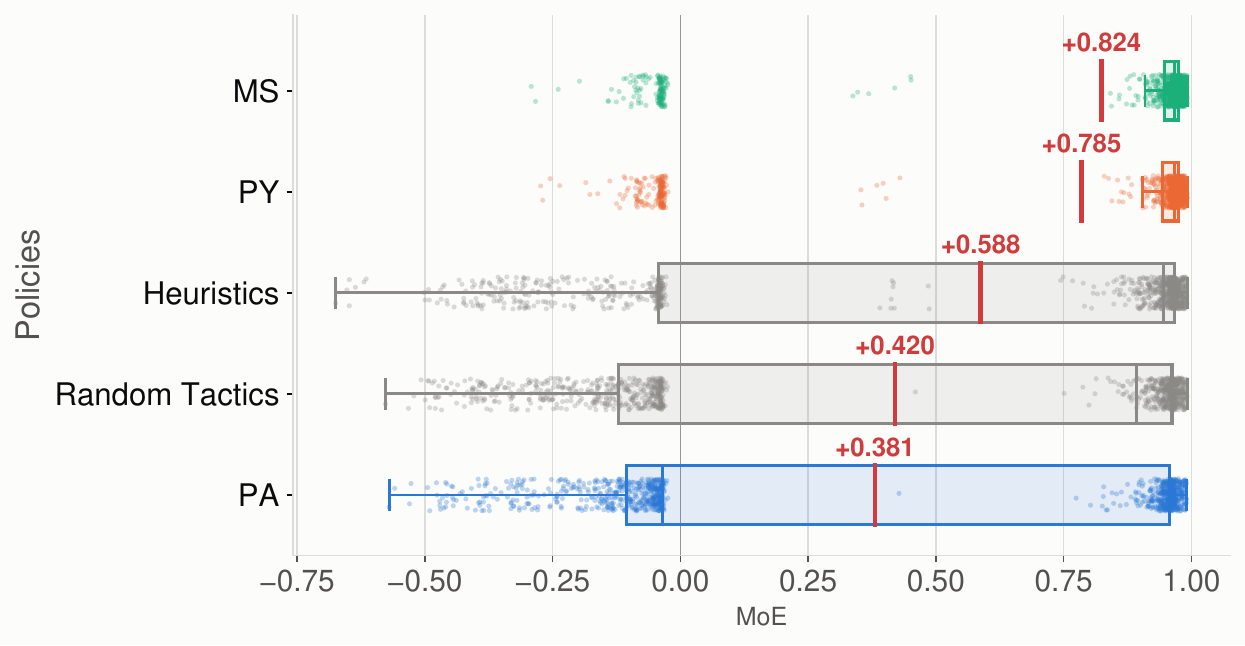}
    \caption{MoE distributions of MS, PY, PA, Heuristics and Random Tactics over 1{,}000 shared episodes in Salamis}
    \label{fig:salamis-eval}
\end{figure}

Even in this unseen scenario, the main findings of RQ2 and RQ3 hold (Figure~\ref{fig:salamis-eval}). MS attains the highest mean MoE (0.824), followed by PY (0.785), the same order as in the pooled evaluation (Figure~\ref{fig:eval_pooled_with_theater}), and both retain a wide margin over Heuristics (0.588) and Random Tactics (0.420). PA, in contrast, is the worst-performing approach: its mean MoE (0.381) falls below those of both baselines. Comparing the three scenarios further reinforces the observation that training transfers from harder to easier scenarios but not the reverse. PY, trained on the hardest scenario, retains its advantage over the baselines in both easier ones, whereas PA, trained on the easiest, falls to or below the level of the baselines in both harder ones. A plausible explanation is that in Palisades, where every policy attains a mean MoE above 0.85, fewer episodes reward a better choice of tactics, whereas in Pyrenees, coordinators are exposed to fires whose containment depends on the choice of tactics.

From an ME perspective, whether a single-scenario coordinator transfers to an unseen scenario depends on its training scenario, and this dependence is difficult to assess before deployment. Multi-scenario training retains its advantage: it attains the highest mean MoE of all five approaches in every scenario, including Salamis, which it never encountered.

\section{Discussion}
\label{sec:discussion}

Traditional SoS architecting, design and evaluation methods struggle answering to changing operational conditions and remain brittle \cite{raz2024ME}. To that extent ME offers a pathway to analyze the dynamic conditions in which an SoS will operate, however the application of advanced analytical tools integrated with high fidelity simulations remains rare in this domain. This study answers the call of \cite{raz2024ME} for adaptive task configuration for ME and employs RL for this purpose. To our knowledge, this is the first ME study to intelligently coordinate missions under operational uncertainties using a combination of high-fidelity digital mission models and RL. The results provide a proof of concept for a fixed set of mission assets and under the restrictive assumption of perfect information.

\paragraph{Efficacy of the approach.} The results confirm that the RL-based intelligent mission coordinator can learn effective firefighting tactics over the course of training. In each scenario, the coordinators trained on that scenario significantly outperformed both the random and the heuristic baselines, whose arbitrary or static tactics cannot adapt to the evolving fire. The nature of this advantage depends on the scenario. In Palisades, where most missions succeed under every policy, the coordinators outperformed the baselines mainly by reducing catastrophic failures, a form of improvement that is particularly relevant for mission-critical systems, where predictability can be as valuable as average performance. In Pyrenees, where most fires escape, the successful coordinators made a difference by containing substantially more (two to three times) fires as the baselines (see Table \ref{tab:containment}).

\paragraph{Single-scenario training can be brittle.}
A coordinator trained on a single scenario may remain brittle when deployed to a new scenario despite outperforming baseline heuristics on its own. The coordinator trained only on Palisades performed no better than Random Tactics in Pyrenees and Salamis and significantly below even the simple heuristic baseline; despite 150{,}000 training episodes, its policy offered no benefit in the new environment. The coordinator trained on the more demanding Pyrenees scenario, by contrast, retained most of its advantage in Palisades and Salamis. We infer that forgiving training environments exert limited pressure to learn tactics that remain effective under harsher conditions, so the choice of training scenario is itself a design decision with consequences for deployment.

\paragraph{Multi-scenario training.}
These findings motivate training a coordinator across multiple scenarios. Within the same simulation budget, the multi-scenario coordinator performed on par with the best single scenario coordinator in each scenario. In the pooled evaluation, it significantly outperformed every other policy, including Pyrenees only trained agent, when both scenarios were considered together. Furthermore, in Salamis, a scenario none of the coordinators saw during training phase, it attained the highest mean MoE. Combining experience from both scenarios therefore did not dilute the coordinator's performance in either, and allowed a single coordinator to attain superior performance and avoid the brittleness of coordinators trained in isolation. From an ME perspective, this suggests that a mission coordinator trained in multiple scenarios gain an advantage over agents trained in isolation. However, we acknowledge multi scenario training's limitations despite its success against coordinators trained in isolation. There are many ways to train across scenarios, and we adopted a simple one, an equal split of the parallel workers. Although this scheme outperformed the two sequential curricula we examined (Section~\ref{sec:curricula}), we do not claim it is optimal, and the curriculum results show that the scheduling of the training budget materially affects the outcome.

\paragraph{Mission uncertainty persists.} Notably, none of the trained coordinators completely eliminated mission failures. Even in Palisades, where the best performing coordinators contained more than 98\% of fires, none avoided every catastrophic event. On the other hand, in Pyrenees, most fires escaped under every policy. We attribute these failures largely to fires that spread beyond what the fixed fleet can suppress under challenging combinations of ignition location and weather; such fires may be uncontainable with the given resources regardless of the tactics employed. Conversely, even the baselines conduct nearly flawless operations in some missions under favorable conditions. Mission outcomes therefore depend strongly on the inherent difficulty of each fire under every policy. Given these outliers in both directions, we stress the importance of rigorous analysis, verification, and validation of learned policies before deployment, since conclusions drawn from limited data can overstate the value of a method. Furthermore, we urge practitioners to audit and continuously monitor AI systems integrated into their missions to anticipate and mitigate undesired outcomes, such as leaving protected populations in firefighting or selecting civilian targets in military missions.

\paragraph{Training variability shapes deployed performance.}
The sensitivity analyses show that brittleness also arises within a single scenario when training does not represent the variability of the operational environment. For instance, an agent trained on a fixed ignition location shows high effectiveness in evaluation under the same settings. However, it proves brittle once the ignition location varies. Similarly, coordinators trained under nominal weather lose absolute performance when deployed to hotter, drier, and windier conditions, although they retain their relative advantage over the baselines (Section~\ref{sec:sens_weather}). Across scenarios, the evaluation in Salamis, which none of the coordinators saw during training, shows that what a coordinator is trained on matters. Every policy attains its lowest mean MoE in Pyrenees and its highest in Palisades, making Pyrenees the hardest of the three scenarios, Salamis of intermediate difficulty, and Palisades the easiest. In Salamis, PA, trained on the easiest scenario, falls below both baselines, whereas PY, trained on the hardest, and MS, trained on both, retain their advantage (Section \ref{sec:unseen-scenario}). Hence, these results indicate that the variability of the threat represented during training, whether across scenarios or across ignition points within a scenario or in environmental conditions, shape the performance of a coordinator after deployment - as expected. We conclude that performance under training conditions alone is not a reliable indicator of operational performance and that evaluation under conditions beyond those seen in training is necessary to uncover brittleness.

\paragraph{From robustness to mission resilience.}
The results concern the robustness of the coordinators across scenarios and conditions, not than their resilience to disruptions during a mission. Mission resilience concerns the ability of the SoS to absorb, adapt to, and recover from disruptions as they occur, such as asset failures, airport closures, and communication delays. The weather and unseen scenario analyses (Sections~\ref{sec:sens_weather} and \ref{sec:unseen-scenario}) offer a first indication that the coordinators' advantage survives unseen conditions, but moving toward mission resilience requires subjecting the coordinators to such perturbations during the mission, and measuring not only the final mission outcome but also how performance degrades and recovers. The dependence of the SoS on its supporting infrastructure, such as airports and water sources, is one of the main concerns as disruptions to this infrastructure directly constrain the operations.

\paragraph{Limitations.} 
The results are subject to several limitations. First, the simulation environment assumes perfect information and complete adherence of assets to orders, conditions that in practice, are often compromised by latency, low visibility, or adversarial interference. Second, we employed a discretized tactical action space (i.e., selection, tracking, and suppression) to facilitate efficient learning. While this abstraction captures high-level tactics, it may preclude the nuanced execution found in continuous control environments. Third, the study used a fixed fleet and did not address asset or fleet selection. This is a fairly large problem that deserves its own dedicated study thus were deemed out of scope. Finally, the heuristic baseline is a single representative static rule rather than a validated model of expert practice nor a dynamically evolving set of heuristics that adapt to the mission condition. Hence our assessment of heuristics might be underestimating their potential value. Nevertheless, while these limitations bound the scope of the conclusions, they do not alter the main conclusions that RL can learn effective tactical coordination in high-fidelity mission models and that multi-scenario training can mitigate the brittleness of single-scenario coordinators.

\section{Conclusions and Future Work}

\subsection{Conclusion}
Because complex engineered systems require long development times, operational needs often evolve before deployment, and ME therefore requires rigorous analytical approaches for asset acquisition, fleet selection, and mission execution; under a vast array of mission and SoS uncertainties that often render traditional tradespace exploration and optimization intractable. As a step toward resilient mission outcomes, this study provides a proof of concept for integrating high-fidelity digital mission models with RL to enable intelligent mission coordination. Trained over repeated simulations, the resulting coordinator policies map the state of the SoS and its environment to tactical actions without explicitly modeling the uncertainty in the mission scenario.


Beyond demonstrating that RL-based mission coordinators can learn effective tactics that outperform baseline approaches in the scenarios they were trained on, this study showed that the conditions represented during training shape a coordinator's performance across operational environments. A coordinator trained on the more forgiving scenario proved brittle when deployed in the more demanding one, whereas a single coordinator trained concurrently across both scenarios matched the specialized coordinators in each and performed best overall, provided that the training budget was scheduled appropriately. Brittleness also persists against changes in threat and environmental characteristics not represented in training. From an ME perspective, the scenarios, threat characteristics, and curricula used in training are therefore design decisions in their own right and should reflect the variability expected in operation. Finally, none of the coordinators eliminated mission failures, which underscores that trained coordinators cannot guarantee appropriate actions in every situation and require rigorous verification and validation before deployment and continuous monitoring throughout their lifecycles. ME serves as a cornerstone for enhancing SoS performance and resilience under environmental uncertainty, and we believe its combination with rigorous AI methods will remain a fruitful research avenue at the intersection of SE and AI.

\subsection{Future Work}
Future research will address real-world constraints such as imperfect information sharing, communication delays, and misinformation. The sensitivity analyses showed that the variability represented during training shapes how well a coordinator generalizes, which motivates extending multi-scenario training to additional scenarios and a wider range of threat characteristics, and, in particular, evaluating coordinators on scenarios held out from training. A related direction is the principled design of multi-scenario training curricula, for instance by adapting the share and ordering of training across scenarios to their difficulty.
This study addressed only the tactical management of a mission executed by a fixed fleet, however ME spans interconnected decision hierarchies across multiple time scales, from mission tactics to portfolio selection, supporting infrastructure design, and the design of new constituent systems. Our long term objective is to leverage this first step to develop methodologies for longer-horizon decisions by evaluating candidate portfolios and designs with the intelligent coordinator in the loop: comparing alternative fleet portfolios, allocating a limited budget of assets across regions, and deriving design requirements for constituent systems, such as the payload and endurance of the eVTOLs. Integrating explainable AI methods, so that trained policies become interpretable and stakeholders gain insight into why particular tactics are chosen, is a further avenue. Finally, to move from robustness to mission resilience, we plan to subject the coordinators to real-time perturbations such as asset failures and airport closures, and to investigate how dependencies on supporting infrastructure influence mission performance and recovery.

\section*{Appendix}
\appendix
\renewcommand{\thesection}{\Alph{section}}
\setcounter{section}{0}

\section{State Variables}

\begin{table}[H]
\centering
\small
\caption{Description of State Variables}
\label{tab:state-space}
\footnotesize
\setlength{\tabcolsep}{2pt}
\begin{tabularx}{\linewidth}{|p{0.28\linewidth}|X|p{0.22\linewidth}|}
\hline
\textbf{Variable Name (Number of Variables)} & \textbf{Description} & \textbf{Unit of Measurement}\\
\hline
Scenario identifier$^{\dagger}$ (3) & Identifies the mission scenario in a one-hot encoded way & Dimensionless\\ \hline
Time since detection & Duration since the detection of the fire & Minutes\\ \hline
Temperature & Current environmental temperature & Celsius degrees\\ \hline
Humidity & Current environmental humidity & Percentage\\ \hline
Wind speed and direction (2) & Current wind speed and direction & m/s and degrees\\ \hline
Time to sunset & Duration until sunset & Minutes\\ \hline
Minimum distance to fire line & Shortest distance from the current fire front to the indirect fire-line segment being built now & Meters\\ \hline
Distance to water & Distance from the fleet to the nearest water resource & Meters\\ \hline
Fire-boundary distances (5) & Shortest distance from the fire boundary to urban locations, water resources, highly combustible areas, high-slope areas, and the indirect fire line & Meters\\ \hline
Ignition point (2) & Location of the episode's ignition & X, Y coordinates\\ \hline
Fire center (2) & Centroid of all burning cells in the map & X, Y coordinates\\ \hline
Fire extreme points (8) & Uppermost, lowermost, leftmost and rightmost coordinates of all burning cells & X, Y coordinates\\ \hline
Fire spread angle & Propagation angle of the fastest-spreading cell relative to the centroid & Degrees\\ \hline
Spread ray hit point (2) & Coordinates of the point where spread direction ray (in the direction of propagation from the fire center) hits the map boundaries & X, Y coordinates\\ \hline
Front severity (3) & Per-sector severity of the fire front, for each of the 3 directional sectors around the fire center & Dimensionless\\ \hline
Asset-protection urgency scores (15) & Per-sector urgency of defending urban (VIP), highly combustible, high-slope, water resources and indirect-line objectives threatened by that front & Dimensionless\\ \hline
Aircraft locations (20) & Current location of each controlled aircraft & X, Y coordinates\\ \hline
\end{tabularx}
\end{table}

\section{Tactics}
\label{sec:app-tactics}

\begin{figure}[H]
  \centering
  \makebox[\textwidth][c]{%
    \includegraphics[width=1.2\textwidth, trim=60 250 60 250, clip]{"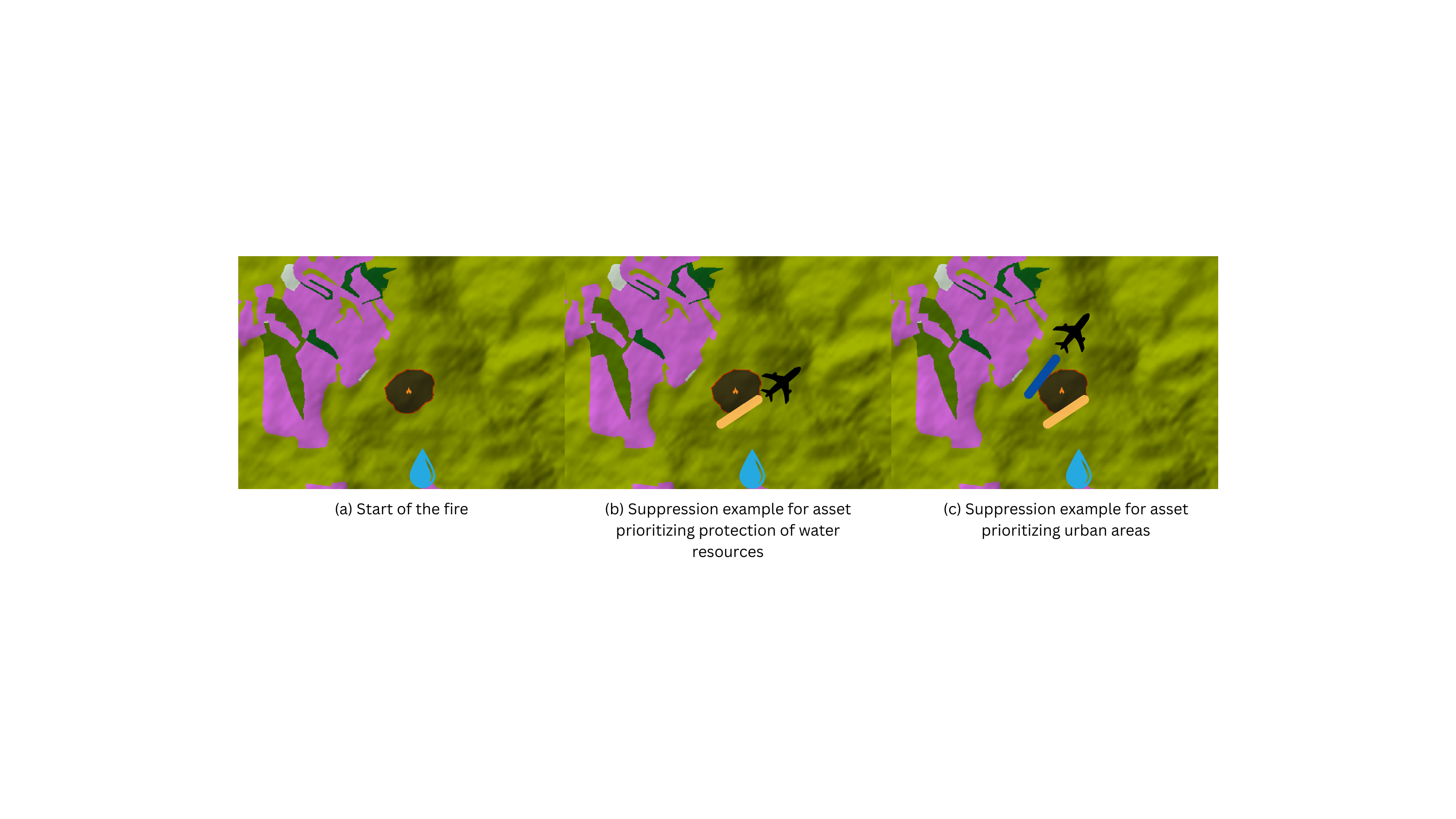"}%
  }
  \caption{Simulation progression, and two example track and suppress operations}
  \label{fig:tactics}
\end{figure}

\indent To visualize these POI decisions, we present Figure \ref{fig:tactics}. Figure \ref{fig:tactics}.a illustrates the moment that the fire is first detected; the green area shows the vegetation, the black represents the area that is already burnt, the red shows the line of fire, and the blue water drop illustrates the relative location of water resources on the map. Figures \ref{fig:tactics}.b and \ref{fig:tactics}.c show two Track POI options, “prioritize water resources” and “prioritize urban areas” with direct suppression. These track POI options are indicated with blue and green suppression actions shown in lines, respectively. Figure \ref{fig:tactics} also allows us to illustrate the complexity and challenge of intelligent mission coordination. As each aircraft takes actions according to the tactic they are assigned, the propagation of the fire is affected by these tactical decisions equivalent to the surrounding conditions at the time such as wind, temperature, and humidity; and the type of cell the fire is spreading to (e.g., highly flammable vegetation vs. urban areas). Collectively, these influence the goodness of mission outcomes as represented by the Measure of Effectiveness (MoE) which is discussed next.

\section{MoE Normalization Constants}

\begin{table}[H]
\caption{Per-scenario normalization constants of the MoE
(Eq.~(3)): the maximum damage each term can inflict, bounded by the
region's characteristics.}
\label{tab:moe-norms}
\footnotesize
\setlength{\tabcolsep}{1pt}
\centering
\begin{tabular}{l r r r r}
\hline
Scenario $i$ & $\text{Max Burnt Area}_i$ (ha) & $\text{Max Cost}_i$ (M\euro) & $\text{Max Emissions}_i$ (tonnes) & $\text{Max Casualties}_i$ (persons) \\
\hline
Palisades & 9{,}087 & 191{,}106 & 131{,}224     & 300{,}000 \\
Pyrenees  & 9{,}938 & 17{,}509  & 2{,}364{,}064 & 10{,}000  \\
Salamis   & 4{,}146 & 13{,}993  & 714{,}009     & 6{,}000   \\
\bottomrule
\hline
\end{tabular}
\end{table}

\section{PPO Advantage and Critic Losses}
\label{app:ppo}
The advantage estimate $\hat{A}_t$ in Eq.~\ref{eq:eq_2} is computed with generalized advantage estimation~\cite{schulman2015gae} from the critic's temporal-difference errors over a rollout of length $T$:
\begin{equation}
\delta_t = \mathcal{R}(s_t, a_t) + \gamma V_{\phi}(s_{t+1}) - V_{\phi}(s_t), \qquad
\hat{A}_t = \sum_{l=0}^{T-t-1} (\gamma\lambda)^{l}\, \delta_{t+l},
\label{eq:gae}
\end{equation}
where $\lambda \in [0,1]$ balances bias and variance. Actions with positive advantage become more likely under the updated policy and those with negative advantage less likely. The critic is trained by minimizing the squared error between its value estimates and the return targets $\hat{R}_t = \hat{A}_t + V_{\phi_{\mathrm{old}}}(s_t)$:
\begin{equation}
L^{V}(\phi) = \hat{\mathbb{E}}_{t}\!\left[\big(V_{\phi}(s_t) - \hat{R}_t\big)^{2}\right],
\label{eq:critic}
\end{equation}
so that its estimates, and hence the advantages it provides, improve as training progresses.

\renewcommand{\bibfont}{\footnotesize}
\bibliographystyle{ama}
\bibliography{FFSoS}
\end{document}